%% file: main_v0.tex
\documentclass[conference]{IEEEtran}

\usepackage{ifpdf}
\usepackage{url}
\ifpdf
\else
\fi

\usepackage{cite}
\ifCLASSINFOpdf
  \usepackage[pdftex]{graphicx}
  \graphicspath{{../pdf/}{../jpeg/}}
  \DeclareGraphicsExtensions{.pdf,.jpeg,.png}
\else
  \usepackage[dvips]{graphicx}
  \graphicspath{{../eps/}}
  \DeclareGraphicsExtensions{.eps}
\fi
\usepackage{amsmath}
\usepackage{algpseudocode}% http://ctan.org/pkg/algorithmicx
\algtext*{EndWhile}% Remove "end while" text
\algtext*{EndIf}% Remove "end if" text
\algtext*{EndFor}
\usepackage{algorithm}
\usepackage{array}
\usepackage{amsfonts}
\usepackage{units}
\usepackage{amssymb}
\usepackage{balance}
\usepackage{tikz} % for tikz
\usepackage[utf8]{inputenc}
\usepackage{pgfplots} 
\usepackage{pgfgantt}
\usepackage{pdflscape}

\usepackage{color}
\newcommand{\complexset}[2]{ \mathbb{C}^{#1 \times #2}  }
\newcommand{\Imatrix}{{ \boldsymbol{\mathrm{I}} }}

\usepackage{acro}
\include{Acronyms}

\usepackage{tabu,longtable}

\ifCLASSOPTIONcompsoc
 \usepackage[caption=false,font=normalsize,labelfont=sf,textfont=sf]{subfig}
\else
 \usepackage[caption=false,font=footnotesize]{subfig}
\fi
\usepackage{stfloats}
\usepackage[hidelinks]{hyperref}
\usepackage{xcolor}
\include{macros}

\begin{document}

% \bstctlcite{IEEEexample:BSTcontrol}
% \title{How Far is the Far-Field? MCRB-based Analysis from a Localization Perspective}
% \title{Far-field and Near-field Model Mismatch Analysis from a Localization Perspective}
\title{Channel Model Mismatch Analysis for XL-MIMO Systems from a Localization Perspective}

\author{Hui Chen\IEEEauthorrefmark{1}, 
Ahmed Elzanaty\IEEEauthorrefmark{2}, 
Reza Ghazalian\IEEEauthorrefmark{3}, 
Musa Furkan Keskin\IEEEauthorrefmark{1},
Riku Jäntti\IEEEauthorrefmark{3},
Henk Wymeersch\IEEEauthorrefmark{1}\\
\IEEEauthorrefmark{1}Chalmers University of Technology, Sweden (\{hui.chen, furkan, henkw\}@chalmers.se)\\
\IEEEauthorrefmark{2}University of Surrey, UK (a.elzanaty@surrey.ac.uk)\\
\IEEEauthorrefmark{3}Aalto University, Finland (\{reza.ghazalian, riku.jantti\}@aalto.fi)
% e-mail: hui.chen@chalmers.se
}

% \title{How Far is the Far-Field? MCRB-based Analysis from a Localization Perspective}
% \title{Determine the Localization Far-field Using MCRB}

% \author{Hui Chen, Tarig Ballal, and Tareq Y. Al-Naffouri
% % Michael~Shell,~\IEEEmembership{Member,~IEEE,}
% %         John~Doe,~\IEEEmembership{Fellow,~OSA,}
%         % and~Jane~Doe,~\IEEEmembership{Life~Fellow,~IEEE}% <-this % stops a space
% \thanks{The authors are with the Division of Computer, Electrical and Mathematical Science \& Engineering, King Abdullah University of Science and Technology (KAUST), Thuwal, 23955-6900, KSA. e-mail: (\{hui.chen; tarig.ahmed; tareq.alnaffouri\}@kaust.edu.sa).}}% <-this % stops a space
% \thanks{J. Doe and J. Doe are with Anonymous University.}% <-this % stops a space
% \IEEEoverridecommandlockouts
% \IEEEpubid{\makebox[\columnwidth]{978-1-5386-5541-2/18/\$31.00~\copyright2018 IEEE \hfill}
% \hspace{\columnsep}\makebox[\columnwidth]{ }}
\maketitle
% \IEEEpubidadjcol

\begin{abstract}
Radio localization is applied in high-frequency (e.g., mmWave and THz) systems to support communication and to provide location-based services without extra infrastructure. {For solving localization problems, a simplified, stationary, narrowband far-field channel model is widely used due to its compact formulation.} However, with increased array size in extra-large MIMO systems and increased bandwidth at upper mmWave bands, the effect of channel spatial non-stationarity (SNS), spherical wave model (SWM), and beam squint effect (BSE) cannot be ignored. 
In this case, localization performance will be affected when an inaccurate channel model deviating from the true model is adopted.
In this work, we employ the MCRB (misspecified Cram\'er-Rao lower bound) to lower bound the localization error using a simplified mismatched model while the observed data is governed by a more complex true model. 
The simulation results show that among all the model impairments, the SNS has the least contribution, the SWM dominates when the distance is small compared to the array size, and the BSE has a more significant effect when the distance is much larger than the array size.
\end{abstract}
% The Frauhofer distance is a rough boundary for near and far field, which is inaccurate and does not consider all the factors (e.g., orientation, array shape).
% (considering all the mismatches). 
% Based on the MCRB, we define the model mismatch boundary as the $\unit[-3]{dB}$ loss of the mismatched lower bound with CRB of the true model.


\IEEEpeerreviewmaketitle

\section{Introduction}
\label{sec:intro}

Radio localization is playing an important role in the fifth/sixth generation (5G/6G) communication systems to support various emerging applications, e.g., autonomous driving~\cite{bresson2017simultaneous}, digital twins~\cite{tao2018digital}, and augmented reality~\cite{siriwardhana2021survey}. In general, localization starts with channel estimation and channel geometric parameters extraction by assuming a sparse channel consisting of a limited number of paths. From the angle and delay estimation with respect to known anchors, e.g., \acp{bs}, the \ac{ue} position can be estimated. Benefiting from the large bandwidth and array size of mmWave and THz systems, high angular and delay resolution, and hence accurate localization performance is expected~\cite{chen2021tutorial}.

Research on 5G/6G radio localization has drawn significant attention recently and the works range from 2D~\cite{shahmansoori2017position} to 3D~\cite{abu2018error} scenarios, from \ac{nlos}-assisted~\cite{mendrzik2018harnessing} to \ac{ris}-supported localization~\cite{elzanaty2021reconfigurable}. Most of the works consider a stationary, narrowband \ac{ff} model due to its simplicity in algorithm design and performance analysis. This simplified model works well in conventional communication systems with limited bandwidth and antennas. Nevertheless, to combat high path loss with signals at high carrier frequencies, \ac{xlmimo} and large \acp{ris} will be deployed, resulting in \ac{sns} and \ac{swm}, which are considered as the \ac{nf} features~\cite{de2020non,marinello2020antenna,elzanaty2021reconfigurable}. In addition, a much wider bandwidth causes \ac{bse} that makes the simplified model insufficient~\cite{tan2021wideband, tarboush2021teramimo}. Therefore, there is a need to understand to what extent the conventional model holds.
%a more accurate model considering all the mismatches is needed.

Some recent works investigate error bounds and develop localization algorithms in \ac{nf} scenarios considering both SNS and SWM~\cite{guerra2021near, rahal2022constrained}, or only SWM~\cite{rinchi2022compressive}. In particular, tracking with filter evaluations~\cite{guerra2021near}, compressive sensing-based algorithm~\cite{rinchi2022compressive} and constrained RIS profile optimization~\cite{rahal2022constrained} are discussed for NF localization. However, the complexity of the \ac{nf} models precludes the development of scalable algorithms for \ac{xlmimo} systems. Some approximations exist, such as second-order approximation of the \ac{ff} model~\cite{le2019massive}, or adopting \ac{aosa} structures considering \ac{nf} across the \acp{sa} while retaining the \ac{ff} model for each \ac{sa}~\cite{tarboush2021teramimo}, {but the qualities of these approximations on localization are not evaluated}. Compared with the \ac{nf} model, the \ac{bse} is considered less frequently in localization and sensing works than communication~\cite{tan2021wideband}. As a result, the model mismatch by adopting a simplified model is an important factor affecting the localization performance, especially for large bandwidth \ac{xlmimo} systems~\cite{de2020non,marinello2020antenna}, and the level of performance loss caused by such approximations needs to be studied. 

In this work, instead of discussing an accurate propagation channel model, we aim to answer the question: \textit{when is the conventional simplified model sufficient?}
To answer this question, we define a model mismatch boundary with the help of \ac{mcrb}~\cite{fortunati2017performance}. The main contributions of this paper can be summarized as follows:
\begin{itemize}
    \item We formulate a ``\ac{tm}'' considering SNS, SWM, BSE and explain how these three types of impairments (w.r.t. the conventional model) can be removed one by one to obtain the conventional model, which is treated as a ``\ac{mm}''.
    \item We resort to MCRB analysis to formulate the lower bound (LB) of localization using a \ac{mm} in processing the data obtained from the \ac{tm}, and define a \ac{mme} as the absolute difference between the LB and the CRB of the TM, normalized by the CRB of the TM.
    \item We provide extensive numerical results for the derived LB and evaluate the contributions of different types of mismatches for various scenarios to provide
    guidelines on when the conventional simplified model does not considerably affect the localization performance.
    %suggestions on the scenarios using a simple localization model without affecting the performance too much.
\end{itemize}
% Starting by describing the mismatched model (\ac{ff} model) and the true model (\ac{nf} model with ) for localization, we use MCRB to formulate the lower bound of the mismatched estimator using a mismatched model processing the data obtained from true model. The model mismatch boundary Different scenarios are evaluated both in theoretical analysis and simulation. We show that the model mismatch boundary is determined by a lot of factors (e.g., transmission power, target direction) and different types of mismatches contribute differently. The model mismatch boundary provides suggestions on the scenarios using a simple \ac{ff} model without losing the localization performance too much.

\section{System and Signal Model}
In this section, we start with the signal model and describe the considered \ac{mm} and \ac{tm}.
% Then describe the \ac{mle} and \ac{mmle}. 
Consider an uplink system with a \ac{bs} equipped with an $N$-element \ac{ula} estimating the location of a single-antenna \ac{ue}. The center of \ac{bs} is at the origin of the global coordinate system, and each antenna is located at {$\bv_n = [0, {(2n-N-1)\lambda_c}/{4}]^\top$, $1\le n\le N$}, where $\lambda_c$ is the wavelength of the carrier frequency $f_c$. We analyze a simple scenario by assuming the system is synchronized and only LOS path exists. Then, the position of UE $\pv$ can be expressed as $\pv = {\tau}{c}\,[\cos(\vartheta), \sin(\vartheta)]^\top$, where $\vartheta$ is the \ac{aoa}, $\tau$ is the \ac{toa} and $c$ is the speed of light.

\subsection{Signal Model}
Considering $x_{g,k}$ ($|x_{g,k}|^2 = {P}$, where $P$ is the average transmission power) as the transmitted \ac{ofdm} symbol at $g$-th transmission ($1\le g \le \mathcal{G}$) and $k$-th subcarrier ($1\le k\le K$), the observation at the \ac{bs} can be formulated as
\begin{equation}
    \yv_{g,k} = {\Wm_{\text{}g}^\top\hv_kx_{g,k}} + \Wm_{\text{}g}^\top\nv_{g,k},
    \label{eq:ideal_far_field_model}
\end{equation}
where $\Wm_{\text{}g}\in \mathbb{C}^{N\times M}$ is the unitary (i.e., $\Wm_{\text{}g}^\mathsf{H}\Wm_{\text{}g}=\Imatrix_M$, where $\Imatrix_M$ is an $M\times M$ identity matrix) combining matrix at the BS for the $g$-th transmission with $M \leq N$ representing the number of \acp{rfc}, $\hv_k \in \complexset{N}{1}$ is the channel vector at the $k$th subcarrier, which is assumed to be constant during $\mathcal{G}$ transmissions, and $\nv_{g,k} \in \complexset{N}{1}$  denotes the noise component following a complex normal distribution $\nv_{g,k}\sim \mathcal{CN}(\mathbf{0}, \sigma_n^2 \Imatrix_N)$, with $\sigma_n^2=N_0 W$, where $N_0$ is the noise power spectral density (PSD) and $W=K\Delta_f$ is the total bandwidth. {In this work, we will focus on the analysis of digital arrays to elimiate the effect of the combining matrix, as $\Wm_g = \Imatrix_N$ can be removed from~\eqref{eq:ideal_far_field_model}.}

% For a fully-connected array where all the antennas are connected to $M$ \acp{rfc} via a phase-shifter network, xxx. For a digital array, however, and thus can be removed. We consider both array structures to evaluate the effect of model mismatch on localization performance. 
% {We assume traditional phase-shifters are adopted and hence $\Wm_{\text{}g}$ is frequency-independent~\cite{tan2021wideband}} with equal amplitude for each element, i.e., $|W_{n,m}| = 1$. 

\subsection{Mismatched Channel Model}
We consider a widely used channel model as the \ac{mm}. The channel vector for the $k$th subcarrier $\hv_k^{\text{MM}} \in \complexset{N}{1}$ can be formulated using a complex channel gain $\alpha$, a steering vector $\av_\text{}(\vartheta) \in \complexset{N}{1}$, and a delay component $D_k(\tau)$ as
\begin{align}
    \hv_k^{\text{MM}} = &\alpha \av_\text{}(\vartheta) D_k(\tau),
    \label{eq:far_field_channel_model}
    \\
    \alpha = & \rho e^{-j\xi} 
    = \frac{\lambda_c}{4\pi \Vert \pv \Vert} e^{-j\xi},
    \label{eq:far_field_gain}
    \\
    % \av_\text{}(\vartheta) = &[1, e^{j\pi\sin(\vartheta)}, \ldots, e^{j(N-1)\pi\sin(\vartheta)}]^\top,
    % \label{eq:far_field_steering_vector}
    % \\
    \av_\text{}(\vartheta) = &[e^{-j\pi\frac{N-1}{2}\sin(\vartheta)}, \ldots, 1, \ldots, e^{-j\pi\frac{1-N}{2}\sin(\vartheta)}]^\top,
    \label{eq:far_field_steering_vector}
    \\
    D_k(\tau) = & e^{-j 2 \pi (f_c + k \Delta_f) \tau} = D_k(\pv) =  e^{-j\frac{2\pi}{\lambda_k}\Vert \pv \Vert},
    \label{eq:far_field_channel_delay}
\end{align}
where $f_k = c/\lambda_k = f_c + k\Delta_f$ is the frequency of the $k$th subcarrier, $\alpha$ is an unknown complex channel gain during the coherence time determined by the environment and antenna radiation pattern. Note that in this model, the channel gain $\alpha$ is the same for all the antennas and subcarriers, and the steering vector is only determined by the \ac{aoa}. The delay term $D_k$ is usually simplified as $e^{-j 2 \pi k \Delta_f \tau}$ since the constant component $e^{-j 2 \pi f_c\tau}$ can be incorporated into the channel gain.

\subsection{True Channel Model}
We compare the \ac{mm} with a \ac{tm}, which is the standard NF model used in array signal processing~\cite{friedlander2019localization} including the \ac{bse}. The channel vector for the \ac{tm} can then be formulated as
% Some test: the estimator seems to be biased (probably because of the initial point and the GD algorithm?), does it matter?
% The difference between \ac{nf} and FF channel models is in the channel vector in~\eqref{eq:far_field_channel_model}. The phase change caused by delay can no longer be expressed using a steering vector and the amplitudes for different frequencies at different antennas are varying. 
\begin{align}
    \hv_{k}^{\text{TM}} = & \alphav_k(\pv) \odot \dv_k(\pv) D_k(\pv),
    \label{eq:near_field_channel_model}
    \\
    \alpha_{k,n}(\pv) = & \alpha \, c_{k,n}(\pv), \ \ c_{k,n}(\pv)= \frac{\lambda_k \Vert \pv \Vert}{\lambda_c \Vert \pv-\bv_n \Vert},    
    \label{eq:near_field_channel_amplitude}
    \\
    d_{k,n}(\pv) = & e^{-j \frac{2\pi}{\lambda_k} (\Vert \pv -\bv_n \Vert - \Vert \pv \Vert)}.
    \label{eq:near_field_steering_vector}
\end{align}
Here, $\alpha$ is defined in~\eqref{eq:far_field_gain}, $\alpha_{k,n}$ is the $n$th element of the vector $\alphav_k$ indicating the channel nonstationarity, and $d_{k,n}$ is the $n$th entry of the delay vector $\dv_k$. 
In contrast to the FF model formulated in~\eqref{eq:far_field_channel_model}--\eqref{eq:far_field_channel_delay}, we can see that three types of model impairments are introduced in~\eqref{eq:near_field_channel_model}--\eqref{eq:near_field_steering_vector} as follows.
\begin{enumerate}
    \item \textbf{\Ac{sns}}: {Non-stationarities represented by $\alpha_{k,n}$ in \eqref{eq:near_field_channel_amplitude} occur because different regions of the array see different propagation paths when the array is large~\cite{de2020non}, i.e., the distance between the source and the various antennas may be quite different in the \ac{nf}}. Also, when the bandwidth of the system is large, different antennas and subcarriers will have different channel amplitudes, as shown in~\eqref{eq:near_field_channel_amplitude}.
    \item \textbf{\Ac{swm}}: In the NF, the phase delay between different antennas can no longer be formulated using a steering vector. Only when the distance between the transceivers is much larger than the array aperture, the term $(\Vert \pv -\bv_n \Vert - \Vert \pv \Vert)$ can be approximated into $\sin(\vartheta){(2n-N-1)}\lambda_c /4$~\cite{le2019massive}. By further replacing $\lambda_k$ with $\lambda_c$ (ignore BSE),~\eqref{eq:near_field_steering_vector} is identical to~\eqref{eq:far_field_steering_vector}. 
    % FF, the steering vector in~\eqref{eq:far_field_steering_vector} is an approximation of the actual SWM by assuming a \ac{pwm}~\cite{bohagen2009spherical}. When the distance between the transceivers is much larger than the array aperture, $\Vert \pv -\bv_n \Vert - \Vert \pv \Vert \approx \frac{2n-N-1}{4}\lambda_c \sin(\vartheta)$. In most of the NF works, only this type of mismatch is considered~\cite{friedlander2019localization, elzanaty2021reconfigurable}.
    \item \textbf{\Ac{bse}}: The steering vector in~\eqref{eq:far_field_steering_vector} is the same for all the subcarriers (frequency-independent). When the \ac{bse} is considered~\cite{tarboush2021teramimo,chen2021tutorial}, the FF steering vector $\av(\vartheta)$ need to be modified to be frequency-dependent as $\tilde \av_k(\vartheta)$, with each element
    $\tilde \av_{k,n}(\vartheta) = e^{j\pi\frac{2n-N-1}{2}\sin(\vartheta)\frac{\lambda_c}{\lambda_k}}$. This mismatch can be seen from $\lambda_k$ (instead of $\lambda_c$) in~\eqref{eq:near_field_steering_vector}.
\end{enumerate}

{Note that the NF model is parameterized by the UE position $\pv$, while the FF model is parameterized by the signal AOA $\vartheta$ and delay $\tau$}. When clock offset is introduced in the delay (which is a practical assumption as perfect synchronization is quite challenging to achieve and maintain), the NF model can still be sufficient for positioning in an unsynchronized system with a single \ac{bs} by exploiting the wavefront curvature.

Typically, the near-field is considered in the area between Fresnel distance $D_\text{F}$ and Fraunhofer distance $D_\text{N}$~\cite{balanis2016antenna} as
\begin{equation}
    D_\text{N}\triangleq 0.62\sqrt{\frac{R^3}{\lambda_c}}
    <r<
    D_\text{F}\triangleq\frac{2R^2}{\lambda_c},
\end{equation}
where $r$ is the distance between the transceivers, and $R$ is the largest dimension of the array. 
However, the Fraunhofer distance is just a simple rule of thumb calculation for the boundary between the \ac{ff} and \ac{nf}, which does not consider the BSE and other system parameters (e.g., transmission power or AOA). In the rest of this work, we will show that this distance is insufficient to suggest when the MM can be used in practice without performance degradation.

% \subsection{Fraunhofer and Fresnel Distance}
% In general, the near-field is considered in the area between Fresnel distance $D_\text{F}$ and Fraunhofer distance $D_\text{N}$~\cite{balanis2016antenna} as $D_\text{N}0.62\sqrt{\frac{R^3}{\lambda_c}}<r<D_\text{F}\frac{2R^2}{\lambda_c}$, where $r$ is the distance between the transceivers and $D_\text{N}$, $D_\text{F}$ can be calculated as
% % \begin{align}
% %     D_\text{N} = & 0.62\sqrt{\frac{R^3}{\lambda_c}}, \\
% %     D_\text{F} = & \frac{2R^2}{\lambda_c},
% % \end{align}
% where $R$ is the largest dimension of the array. 
% However, the Fraunhofer distance is just a simple rule of thumb calculation for the far and near field boundary, which does not consider the \ac{snr}, direction, communication system type (e.g., MIMO or SIMO). In the rest of this work, we will show that this distance cannot capture all the factors that differentiate NF and FF models. To start with, we briefly describe the estimators and lower bounds.

\subsection{Summary of the Channel Models}
In this section, we described a MM (FF channel model) and a TM (NF model with BSE). To facilitate the analysis in the following, we further define several \acp{tm}, and all the considered models are summarized as follows:
\begin{enumerate}
    \item MM: the model described in~\eqref{eq:far_field_channel_model}--\eqref{eq:far_field_channel_delay} is considered as the MM.
    \item TM: the model involves all the model impairments described in~\eqref{eq:near_field_channel_model}-\eqref{eq:near_field_steering_vector} is considered as the TM.\footnote{Note that the so-called `true model' is not guaranteed to be the correct model in some applications. Other factors such as the effective antenna areas~\cite{bjornson2020power}, electromagnetic propagation model~\cite{friedlander2019localization} and hardware distortion can also be considered and the mismatch analysis could be conducted similarly.}
    %The effective antenna areas~\cite{bjornson2020power}, electromagnetic propagation model~\cite{friedlander2019localization} and, path loss components for different propagation environment~\cite{wu2021intelligent}, can also be considered and the mismatch analysis could be conducted similarly.
    % In other words, the localization performance loss caused by the mismatch between the FF model and the real model is likely to be larger than the mismatch with the NF model (as formulated in this work).
    \item Other TMs: we use TM-SNS, TM-SWM, TM-BSE to indicate the models that only consider SNS, SWM, and BSE, respectively (one type of impairment at a time).
\end{enumerate}

\section{Estimators and Lower Bounds}
In this section, we briefly describe the \ac{mle} and the \ac{mmle} for localization, and derive the \ac{crb} and \acp{lb} (based on \ac{mcrb}) for the two estimators, respectively.
% in~\eqref{eq:MLE} and the MCLB for the mismatched estimator in~\eqref{eq:MMLE}.

\subsection{Maximum Likelihood Estimator}
For an observed signal vector from the TM $\yv \sim f_\text{TM}( \yv|\alpha, \pv)$ (where $\yv\in \mathbb{C}^{\mathcal{G}K\times 1}$ is a concatenation of all the received symbols from different subcarriers and transmissions), 
the \ac{mle} of the UE position and channel gain is %can be estimated by maximizing the likelihood function as
\begin{align}
    [\hat \pv_\text{MLE} ,\hat \alpha_\text{MLE}]
    & = \arg \max_{\pv,\alpha} \ln f_\text{TM}( \yv|\alpha, \pv),
\end{align}
where $\ln f_\text{TM}( \yv|\alpha, \pv)$ is the log-likelihood of the TM.
We can then use a plug-in estimate to remove the nuisance parameter $\alpha$~\cite{chen2021mcrb}. Then, the position estimation can be obtained as 
\begin{align}
      \hat \pv_\text{MLE}& = \arg \min_\pv \left \Vert  \yv - \frac{\bar\etav(\pv)^{\mathsf{H}}  \yv}{\Vert \bar\etav(\pv)\Vert^2}\bar\etav(\pv) \right \Vert^2,
\label{eq:MLE}
\end{align}
where $\bar\etav(\pv) = \bar\muv(\alpha, \pv)/\alpha$, and $\bar\muv(\alpha, \pv) = [\bar\muv_{1,1}^\top, \ldots, \bar\muv_{G,K}^\top]^\top$ is the concatenation of all the noise-free observations of the TM with $\bar\muv_{g,k} = \Wm_{\text{}g}^\top\hv^{\text{TM}}_k x_{g,k}$ based on~\eqref{eq:ideal_far_field_model} and~\eqref{eq:near_field_channel_model}. The problem in~\eqref{eq:MLE} can be solved by gradient descent with backtracking line search~\cite{nocedal2006numerical}. An initial point for the algorithm can be obtained through a 2D coarse grid search.
% as shown in the Algorithm~1.

Considering MM, the \ac{mmle} can be formulated as 
\begin{align}
    \hat \pv_\text{MMLE}& = \arg \min_\pv \left \Vert  \yv - \frac{\tilde \etav(\pv)^{\mathsf{H}}  \yv}{\Vert \tilde \etav(\pv)\Vert^2}\tilde \etav(\pv) \right \Vert^2,
    \label{eq:MMLE}
\end{align}
with $\tilde \etav(\pv)=\tilde \muv(\alpha, \pv)/\alpha$, and $\tilde\muv(\alpha, \pv) = [\tilde\muv_{1,1}^\top, \ldots, \tilde\muv_{G,K}^\top]^\top$ is the concatenation of all the noise-free observations of the MM with $\tilde\muv_{g,k} = \Wm_{\text{}g}^\top\hv^{\text{MM}}_k x_{g,k}$ based on~\eqref{eq:ideal_far_field_model} and~\eqref{eq:far_field_channel_model}.

\subsection{CRB}
\label{sec:crb}
The \ac{crb} is sufficient to lower bound the \ac{mle}. We define a channel parameter vector as ${\boldsymbol\theta_\text{c}} = [\vartheta, \tau, \rho, \xi]^\top$ and a state vector ${\boldsymbol\theta_\text{s}} = [p_x, p_y, \rho, \xi]^\top$. In this work, these two vectors have a one-to-one mapping and either of them can sufficiently describe the channel model. Regardless of which vector is used, the FIM of the vector $\thetav$ can be expressed as~\cite{chen2021tutorial}
\begin{equation}
    {\mathcal{I}}({\boldsymbol\theta})
    = \frac{2}{\sigma_n^2}\sum^{\Gc}_{g=1} \sum^K_{k=1}\mathrm{Re}\left\{
    \left(\frac{\partial\mu_{g,k}}{\partial{\boldsymbol\theta}}\right)^{\mathsf{H}} 
    \left(\frac{\partial\mu_{g,k}}{\partial{\boldsymbol\theta}}\right)\right\},
    \label{eq:FIM}
\end{equation}
where $\mu_{g,k}$ could be $\bar \mu_{g,k}$ or $\tilde \mu_{g,k}$ to derive the CRB for the TM or MM, $\thetav$ could be $\thetav_\text{c}$ or $\thetav_\text{s}$ and the corresponding FIM $\mathcal{I}(\thetav_\text{c})$ or $\mathcal{I}(\thetav_\text{s})$ can be obtained. The derivative of TM can be found in the Appendix~\ref{sec:appendix_A}.
Usually, the FF model starts with the channel parameters since the channel depends on angle and delay as described in~\eqref{eq:MLE}, whereas the NF model calculates the FIM of $\thetav_\text{s}$ directly as $\mathcal{I}(\thetav_\text{s})$. 
However, the scenario in this work is a special case (synchronized and LOS channel only) and the two FIMs can be transformed from each other through a Jacobian matrix $\Jm_\text{c}$ or $\Jm_\text{s}$ as
\begin{align}
    \mathcal{I}(\thetav_\text{s}) 
    & = \Jm_\text{s}\mathcal{I}(\thetav_\text{c})\Jm_\text{s}^\top,  
    \ \ \Jm_\text{s} = 
    \begin{bmatrix}
        \tilde \Jm_\text{s} & \mathbf{0}_2\\
        \mathbf{0}_2 & \Imatrix{I}_2
    \end{bmatrix},\\
    \mathcal{I}(\thetav_\text{c}) 
    & = \Jm_\text{c}\mathcal{I}(\thetav_\text{s})\Jm_\text{c}^\top,  
    \ \ \Jm_\text{c} = 
    \begin{bmatrix}
        \tilde \Jm_\text{c} & \mathbf{0}_2\\
        \mathbf{0}_2 & \Imatrix{I}_2
    \end{bmatrix},
\end{align}
where $\mathbf{0}_2$ is a $2\times 2$ zero matrix.
$\tilde \Jm_\mathrm{s} = [{\partial \vartheta}/{\partial \pv}, {\partial \tau}/{\partial \pv}]$ is the Jacobian matrix from angle/delay to position using a denominator-layout notation with ${\partial \vartheta}/{\partial \pv}={1}/{(c \tau)}[-\sin(\vartheta),\cos(\vartheta)]^\top$ and ${\partial \tau}/{\partial \pv}={\pv}/{(c \tau)}$. 
Similarly, $\tilde \Jm_\mathrm{c} = [({\partial \pv}/{\partial \vartheta})^\top, ({\partial \pv^\top}/{\partial \tau})^\top]$ is the Jacobian matrix from position to angle/delay to position as ${\partial \pv}/{\partial \vartheta}={c \tau}[-\sin(\vartheta),\cos(\vartheta)]^\top$ 
and ${\partial \pv}/{\partial \tau}=c[\cos(\vartheta), \sin(\vartheta)]^\top$.

Based on the above discussions, we can define the position error bound (PEB), angle error bound (AEB), and delay error bound (DEB) as
\begin{align}
\mathrm{PEB} & = \sqrt{\trace(\mathcal{I}({\boldsymbol\theta_\text{s}})^{-1})},
\label{eq:PEB}\\
\mathrm{AEB} & = \sqrt{([\mathcal{I}({\boldsymbol\theta_\text{c}})^{-1}]_{1, 1})}
\label{eq:AEB},\\
\mathrm{DEB} & = \sqrt{([\mathcal{I}({\boldsymbol\theta_\text{c}})^{-1}]_{2, 2})}
\label{eq:DEB},
\end{align}
where $\trace(\cdot)$ is the trace operation, and $[\cdot]_{i,j}$ is getting the element in the $i$th row, $j$th column of a matrix. The bounds from~\eqref{eq:PEB}--\eqref{eq:DEB} can assist us to evaluate the position, angle, and delay estimation performance that will be affected by the model mismatch.

\subsection{MCRB}
The CRB described in Section~\ref{sec:crb} can be implemented for performance analysis when the models used for the estimator and the data generation are the same. When a mismatched model is implemented, we need to rely on MCRB to analyze the performance.
The \ac{lb} of a mismatched estimator can be obtained as~\cite{fortunati2017performance}
\begin{align}
    \text{LB} = \text{LB}(\bar {\boldsymbol\theta}, {\boldsymbol\theta}_0)
    %& = \text{MCRB}({\boldsymbol\theta}_0) + \text{Bias}({\boldsymbol\theta}_0) \\
    & = \underbrace{\Am_{{\boldsymbol\theta}_0}^{-1}\Bm_{{\boldsymbol\theta}_0}\Am_{{\boldsymbol\theta}_0}^{-1}}_{\text{MCRB}({\boldsymbol\theta}_0)} + \underbrace{(\bar{\boldsymbol\theta} - {\boldsymbol\theta}_0)(\bar{\boldsymbol\theta} - {\boldsymbol\theta}_0)^\top}_{\text{Bias}({\boldsymbol\theta}_0)}, 
    \label{eq:LB_MCRB}
\end{align}
where $\bar \thetav$ is the parameter vector for the \ac{tm} and $\thetav_0$ is the pseudo-true parameter vector by minimizing the KL divergence between $f_\text{TM}(\yv|\bar {\boldsymbol\theta})$ and $f_\text{MM}(\yv| {\boldsymbol\theta})$, $\Am$ and $\Bm$ are two possible generalizations of the FIMs as~\cite{chen2021mcrb}
\begin{align}
    {\boldsymbol\theta}_0 
    & = \arg \min_{{\boldsymbol\theta}} \Vert \bar\muv(\bar {\boldsymbol\theta}) - \tilde\muv({\boldsymbol\theta})\Vert^2,
    \\
    [\Am_{{\boldsymbol\theta}_0}]_{i,j}
    & =  \left. \frac{2}{\sigma_n^2}\text{Re}\left[\frac{\partial^2\tilde\muv({\boldsymbol\theta})}{\partial \theta_i \partial \theta_j}\epsilonv({\boldsymbol\theta}) -  \frac{\partial\tilde\muv({\boldsymbol\theta})}{\partial \theta_j}
    \left(\frac{\partial\tilde\muv({\boldsymbol\theta})}{\partial \theta_i} \right)^{\mathsf{H}}\right]
    \right|_{{\boldsymbol\theta} = {\boldsymbol\theta}_0},
    \\
    \label{eq:matrix_A}
    [\Bm_{{\boldsymbol\theta}_0} ]_{i,j} 
    & =  
    \frac{4}{\sigma_n^4}
        \text{Re} \left[
        \frac{\partial^2\tilde\muv({\boldsymbol\theta})}{\partial \theta_i}\epsilonv({\boldsymbol\theta})
        \right]
        \text{Re} \left[
        \frac{\partial^2\tilde\muv({\boldsymbol\theta})}{\partial \theta_j}\epsilonv({\boldsymbol\theta})
        \right] \notag  \\
        & \left. + 
        \frac{2}{\sigma_n^2}\text{Re}\left[  \frac{\partial\tilde\muv({\boldsymbol\theta})}{\partial \theta_j}
        \left(\frac{\partial\tilde\muv({\boldsymbol\theta})}{\partial \theta_i} \right)^{\mathsf{H}}
        \right]
        \right|_{{\boldsymbol\theta} = {\boldsymbol\theta}_0}.    
\end{align}
Here, $\epsilonv(\thetav) \triangleq \bar\muv(\bar {\boldsymbol\theta}) - \muv({\boldsymbol\theta})$ and the detailed derivation can be found in~\cite{chen2021mcrb}.
The obtained LB satisfies
\begin{align}
    \mathbb{E}_\text{TM} \{ (\hat{\boldsymbol{\theta}}_{\text{MMLE}} -\bar{\thetav})(\hat{\boldsymbol{\theta}}_{\text{MMLE}} -\bar{\thetav})^\top \} \succeq \text{LB}(\bar {\boldsymbol\theta}, {\boldsymbol\theta}_0),
\end{align}
which can be used to evaluate the performance of a mismatched estimator.

\subsection{Model Mismatch Boundary}
Based on the LB derived from previous sections, we can define a \ac{mme} as the log-normalized difference between the LB and the CRB of the true model as
% \begin{equation}
%     E(\thetav) = 10\log_{10}(\frac{|\text{CRB}_\text{TM}(\thetav)-\text{{LB}}(\thetav)|}{\text{CRB}_\text{TM}(\thetav)})
% \end{equation}
\begin{equation}
    \text{MME} = 10\log_{10}\left(\frac{|\text{CRB}_\text{TM}-\text{{LB}}|}{\text{CRB}_\text{TM}}\right).
\end{equation}
Here, $\text{MME}$, $\text{CRB}_\text{TM}$ and $\text{{LB}}$ are general terms could be used to indicate angle, delay or position estimation performance.
We further define a model mismatch boundary between the TM and the MM as the $\unit[-3]{dB}$ contour of the MME. 
Although there is no closed-form solution for this boundary, it is helpful to identify in which area we can use a simplified mismatched model.

\section{Simulation}
\subsection{Simulation Parameters}
We consider a BS with $N = 64$ antennas and set default parameters as follows: average transmission power $P = \unit[20]{dBm}$, carrier frequency $f_c = \unit[140]{GHz}$, bandwidth $W = \unit[400]{MHz}$, number of transmissions $\mathcal{G} = 1/50$ (digital/analog), number of subcarriers $K = 10$, noise PSD $N_0 = \unit[-173.855]{dBm/Hz}$ and noise figure $N_f = \unit[10]{dBm}$.
Matlab code is available at~\cite{chenhui07c8}.
% \href{https://github.com/chenhui07c8/Radio_Localization}{[Github]}.

% https://github.com/chenhui07c8

\subsection{Estimators vs. Lower Bounds}
We first evaluate the MLE and MMLE estimators described in~\eqref{eq:MLE} and~\eqref{eq:MMLE}, and compare the performance with several bounds, namely, CRB of the TM (CRB-TM), CRB of the MM (CRB-MM), and LB of the mismatched model. The UE is located at $\pv = [2, 2]^\top$ and $500$ simulations are performed for each points. From the figure, we found that CRB-TM and CRB-MM are similar for this scenario. In addition, the LB saturates at a certain level of transmission power. This is because for a large $P$, the $\text{MCRB}(\thetav_0)$ in~\eqref{eq:LB_MCRB} is close to zero and only the $\text{Bias}(\thetav_0)$ is contributing to the LB. The derived LB aligns well with the estimator (MMLE), verifying the effectiveness of using MCRB as an analysis tool.

\begin{figure}[t!]
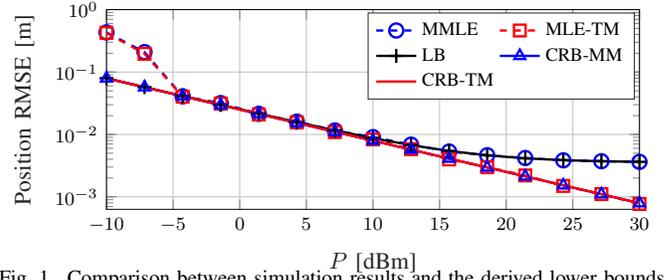

% \begin{minipage}[b]{0.78\linewidth}
    \centering
%   \centerline{\includegraphics[width=0.8\linewidth]{Figures/Fig-1.eps}}
    \include{Figures_tikz/Fig-1}
% \end{minipage}
\vspace{-1cm}
\caption{Comparison between simulation results and the derived lower bounds (LB, CRB-TM, and CRB-MM). The proposed localization algorithm attaches the bound when the average transmission power $P$ is greater than $\unit[-5]{dBm}$, and the LB diverges from the CRBs when $P>\unit[10]{dBm}$.}
\label{fig:estimator_bounds_comparison}
\end{figure}

\begin{figure}[t]
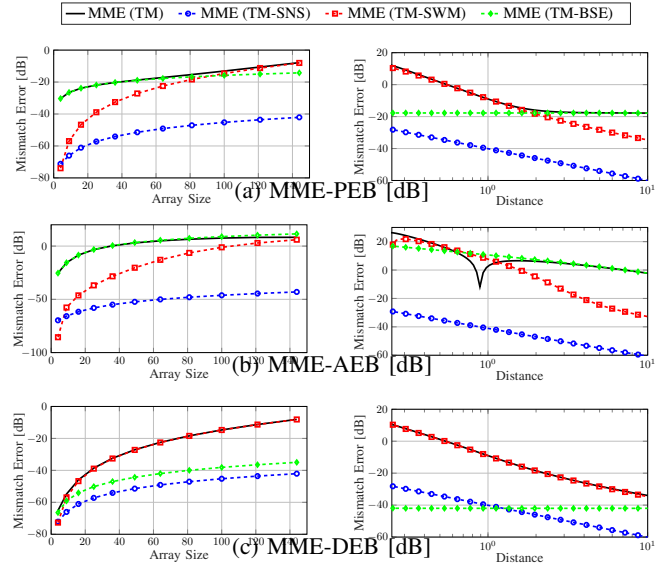

\hspace{0.1mm}
\begin{minipage}[b]{0.48\linewidth}
    \centering
%   \centerline{\includegraphics[width=0.98\linewidth]{Figures/Fig-5-1-a.eps}}
    \include{Figures_tikz/Fig-5-1-a}
    \vspace{-1cm}
\end{minipage}
\begin{minipage}[b]{0.48\linewidth}
  \centering
%   \centerline{\includegraphics[width=0.98\linewidth]{Figures/Fig-5-2-a.eps}}
    \include{Figures_tikz/Fig-5-2-a}
    \vspace{-1cm}
\end{minipage}
\vspace{0.3cm}
\centerline{\small{(a) MME-PEB [dB]}} 
% \medskip
\begin{minipage}[b]{0.48\linewidth}
  \centering
%   \centerline{\includegraphics[width=0.98\linewidth]{Figures/Fig-5-1-b.eps}}
    \include{Figures_tikz/Fig-5-1-b}
    \vspace{-1cm}
\end{minipage}
\begin{minipage}[b]{0.48\linewidth}
  \centering
%   \centerline{\includegraphics[width=0.98\linewidth]{Figures/Fig-5-2-b.eps}}
    \include{Figures_tikz/Fig-5-2-b}
    \vspace{-1cm}
\end{minipage}
\vspace{0.3cm}
\centerline{\small{(b) MME-AEB [dB]}} 
\begin{minipage}[b]{0.48\linewidth}
    \centering
%   \centerline{\includegraphics[width=0.98\linewidth]{Figures/Fig-5-1-c.eps}}
    \include{Figures_tikz/Fig-5-1-c}
    \vspace{-1cm}
    % \centerline{(b) CRB-FF}\medskip
\end{minipage}
\begin{minipage}[b]{0.48\linewidth}
  \centering
%   \centerline{\includegraphics[width=0.98\linewidth]{Figures/Fig-5-2-c.eps}}
    \include{Figures_tikz/Fig-5-2-c}
    \vspace{-1cm}
    % \centerline{(a) LB} \medskip
\end{minipage}
\centerline{\small{(c) MME-DEB [dB]}} 
\vspace{-0.5cm}
\caption{Evaluation of different types of MMEs in terms of (a) MME-PEB, (b) MME-AEB, (c) MME-DEB for varying array size (figures in the left column) and varying distance (figures in the right column). We notice that the effect of SNS dominates in the far-field scenario (small array size and large distance), the effect of SWM dominates in the near-field scenario, and the SNS has the least contribution among the three types of model impairments.}
\label{fig:macro_scale}
\end{figure}

\subsection{Evaluation of Different Types of Impairments}
We evaluate the effect of three types of impairments considered in the \ac{tm}, namely, SNS, SWM and BSE for different array size (4 to 144) and different distance ($\unit[0.25]{m}$ to $\unit[10]{m}$, with Fresnel distance $D_N = \unit[0.235]{m}$ and Fraunhofer distance $D_F=\unit[4.25]{m}$). The bandwidth is chosen as $W = \unit[100]{MHz}$ to reduce the effect of BSE. By considering the mismatches one at a time\footnote{A more reasonable approach is to exclude the impairments one by one in the TM to create new MMs, however, MCRB needs to be derived for each of the new MM. For convenience, we consider different types of mismatches independently.}, we can see different types of relative mismatch errors as shown in Fig.~\ref{fig:macro_scale}. We can see that the effect of BSE dominates in the far-field scenario (small array size and large distance) and has a larger effect on angle estimation. The effect of SWM dominates in the near-field scenario and affects delay estimation more. The SNS has the least contribution in the model mismatch. {When in the far-field, the effects of SWM and SNS are expected to disappear (since SWM and SNS are near-field features), then the BSE dominates as it is the only impairment left. Considering the BSE changes the beam pattern of the array while the SWM changes the phases on the received symbols, these two impairments affect angle and delay estimations differently.} 

\begin{figure*}[htb]
\begin{minipage}[b]{0.325\linewidth}
  \centering
  \centerline{\includegraphics[width=0.98\linewidth]{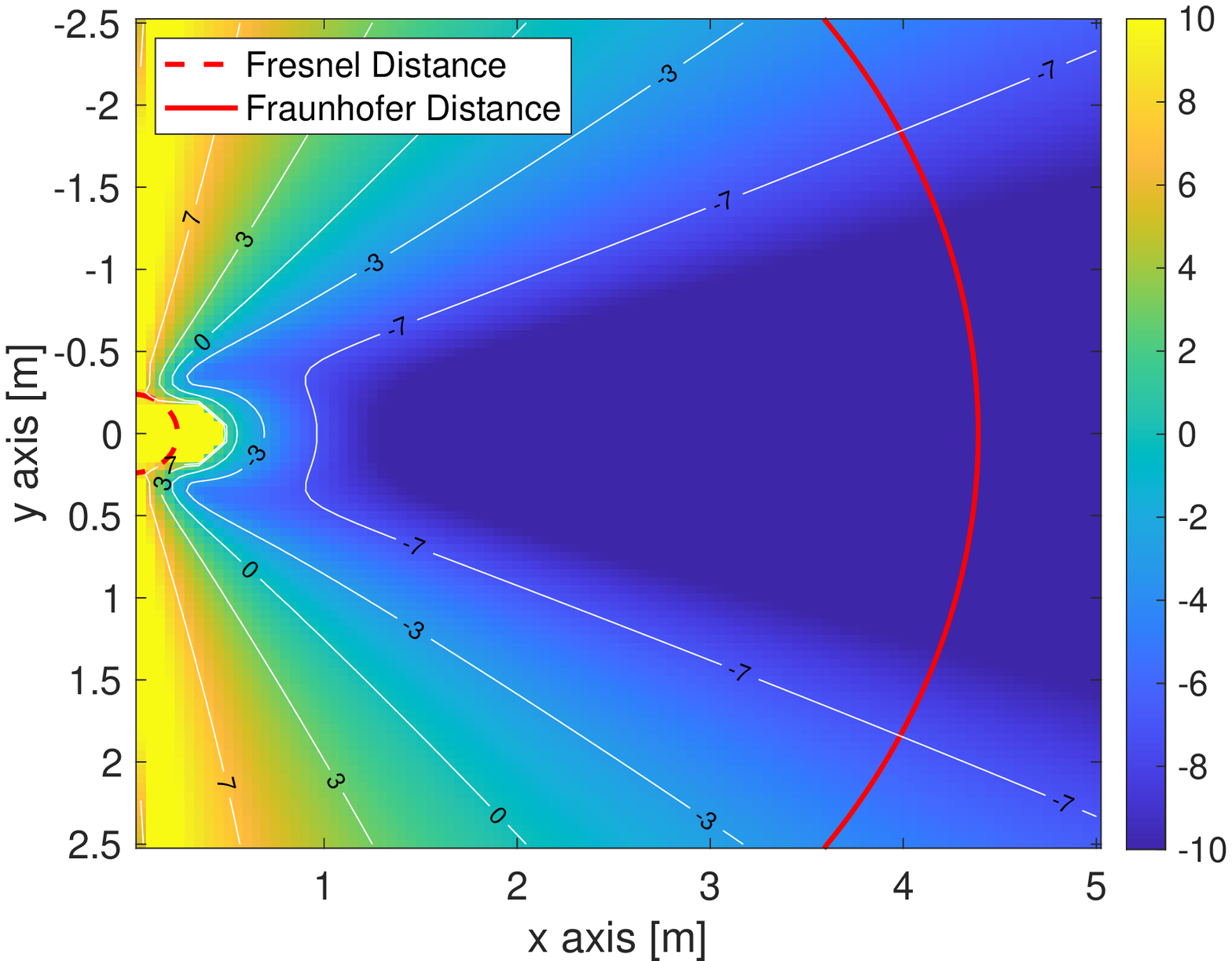}}
    % \include{Figures/fig-2-1}
    % \vspace{-1cm}
    \centerline{\small{(a) MME-PEB [dB]}} \medskip
\end{minipage}
\hfill
\begin{minipage}[b]{0.325\linewidth}
  \centering
  \centerline{\includegraphics[width=0.98\linewidth]{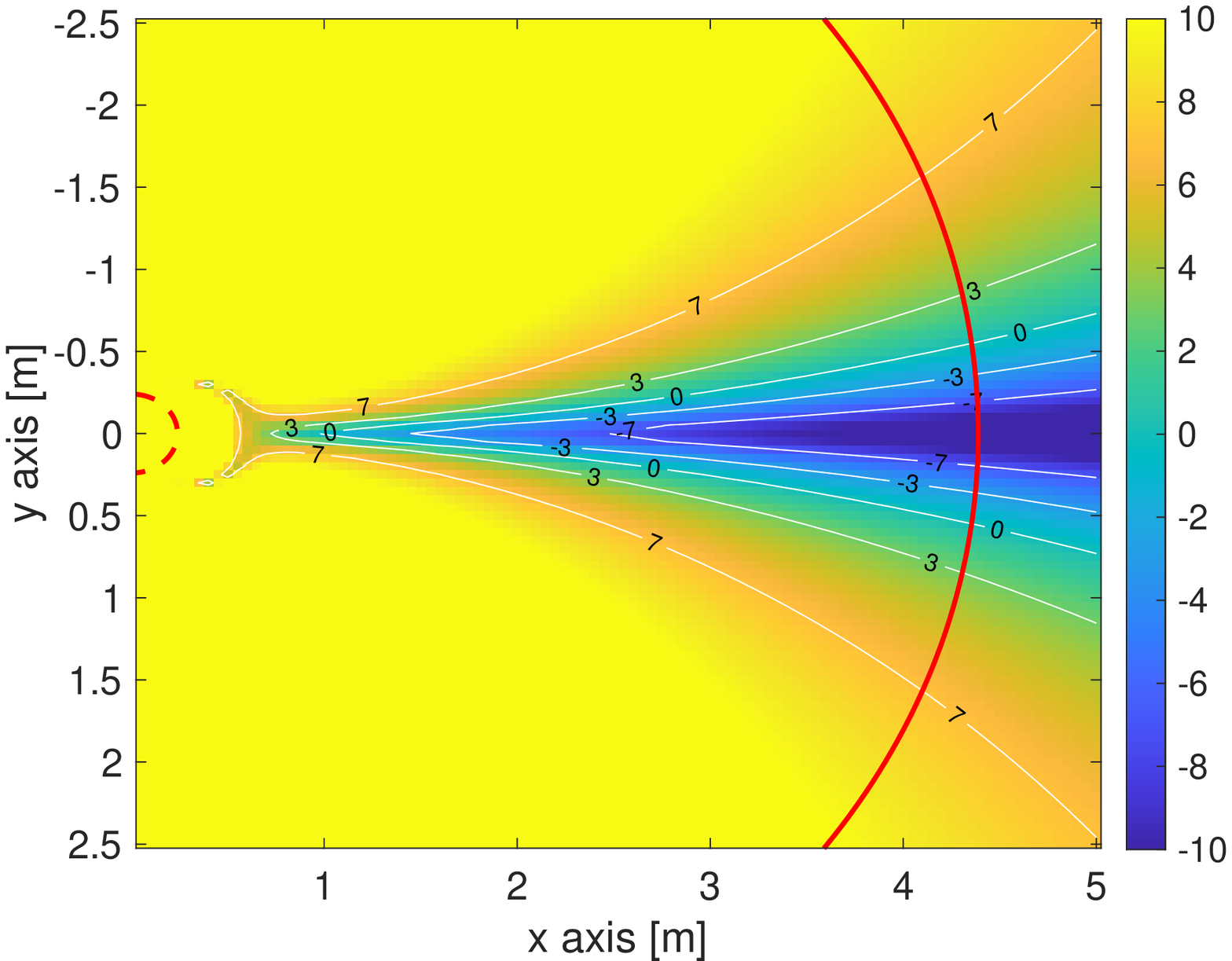}}
    % \include{Figures/fig-2-2}
    % \vspace{-1cm}
    \centerline{\small{(b) MME-AEB [dB]}}\medskip
\end{minipage}
\begin{minipage}[b]{0.325\linewidth}
  \centering
  \centerline{\includegraphics[width=0.98\linewidth]{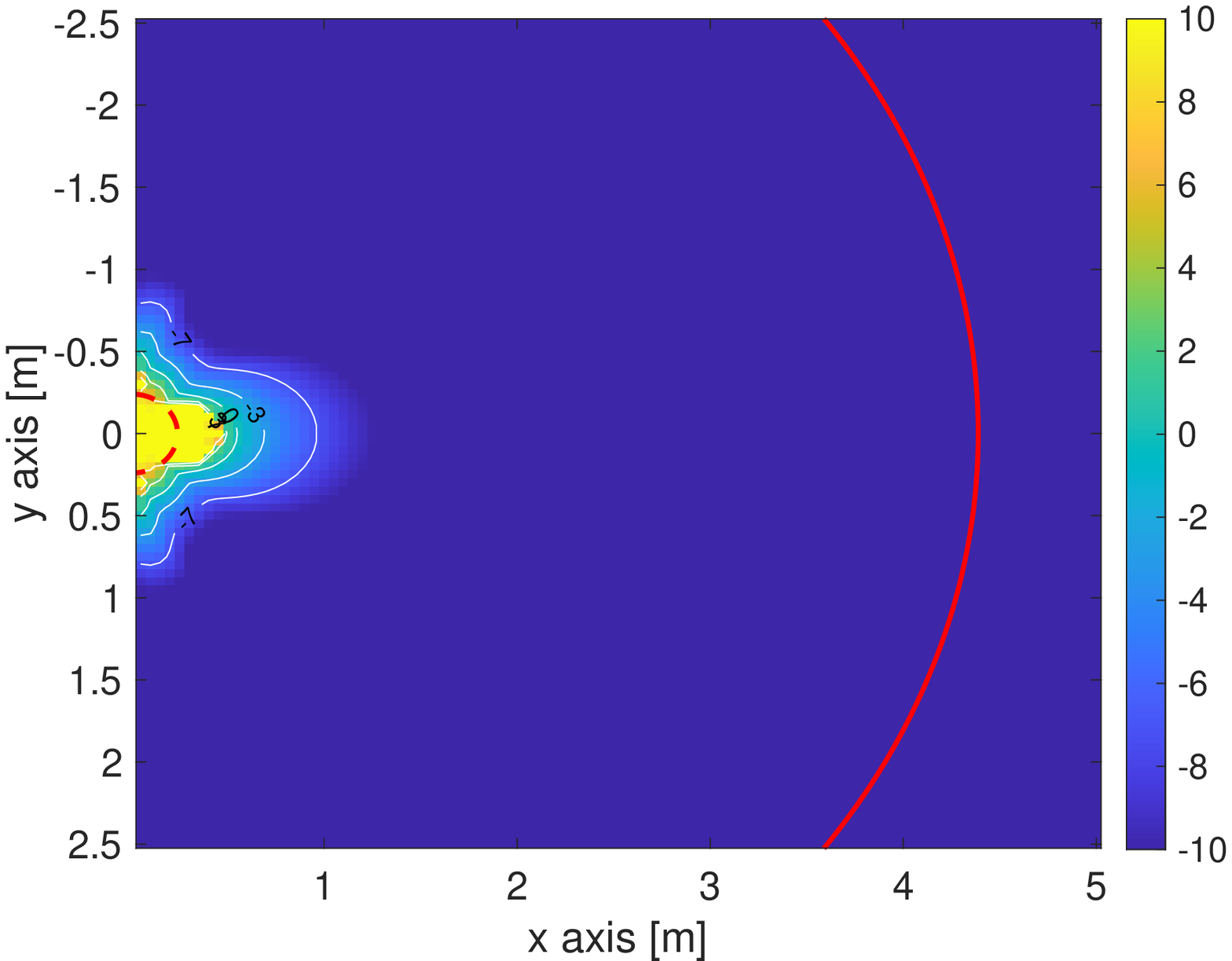}}
    % \include{Figures/fig-2-1}
    % \vspace{-1cm}
    \centerline{\small{(c) MME-DEB [dB]}} \medskip
\end{minipage}
\vspace{-0.5cm}
\caption{Visualization of the model mismatch errors MME-PEB, MME-AEB, and MME-DEB for different UE positions (Digital, $G=1$, $P=\unit[20]{dBm}$, $N=64$). We can see that the angle estimation is more affected than delay estimation due to model mismatch.}
\label{fig:micro_scale}
\end{figure*}

{We notice that there exists a sudden jump at around $\unit[0.9]{m}$ in Fig.~\ref{fig:macro_scale} (b), which can also be seen in Fig.~\ref{fig:micro_scale} (b). This finding indicates that angle estimation error does not contribute to positioning error a lot when the distance is small. In addition, it shows that the model mismatch does not necessarily lead to degradation in localization performance
as the black curve (three types of model impairments) is lower than the blue and green curves (one type of model impairment), which can also be observed in the MCRB analysis considering hardware imperfections~\cite{chen2021mcrb}.}
%the localization performance could be better with model mismatch 

% \subsection{PEB-LB in Macro Scale}
% We show the three bounds in a large space (i.e., $\unit[50\times50]{m^2}$). From (a) and (b) we can see that CRB-PWM is similar to CRB-SWM, which is reasonable since the two models are similar in a far distance, and the performance depends on the SNR, which is reflected by distance (with a fixed transmission power). However, if a mismatched model is used, LB will be different from the true model (SWM), especially in the area where the distance is small.

% \begin{figure*}[t]
% \begin{minipage}[b]{0.325\linewidth}
%   \centering
%   \centerline{\includegraphics[width=0.98\linewidth]{Figures/Fig-2-1.eps}}
%     % \include{Figures/fig-2-1}
%     % \vspace{-1cm}
%     \centerline{(a) CRB-NF} \medskip
% \end{minipage}
% \hfill
% \begin{minipage}[b]{0.325\linewidth}
%   \centering
%   \centerline{\includegraphics[width=0.98\linewidth]{Figures/Fig-2-2.eps}}
%     % \include{Figures/fig-2-2}
%     % \vspace{-1cm}
%     \centerline{(b) CRB-FF}\medskip
% \end{minipage}
% \begin{minipage}[b]{0.325\linewidth}
%   \centering
%   \centerline{\includegraphics[width=0.98\linewidth]{Figures/Fig-2-3.eps}}
%     % \include{Figures/fig-2-1}
%     % \vspace{-1cm}
%     \centerline{(a) LB} \medskip
% \end{minipage}
% \vspace{-0.5cm}
% \caption{Different types of bounds for different UE positions. (a) CRB-NF, (b) CRB-FF, and (c) LB. Model mismatch affects the localization performance, especially when the transceivers are close to each other.}
% \label{fig:macro_scale}
% \end{figure*}

\begin{figure*}[t]
\begin{minipage}[b]{0.245\linewidth}
  \centering
  \centerline{\includegraphics[width=1.1\linewidth]{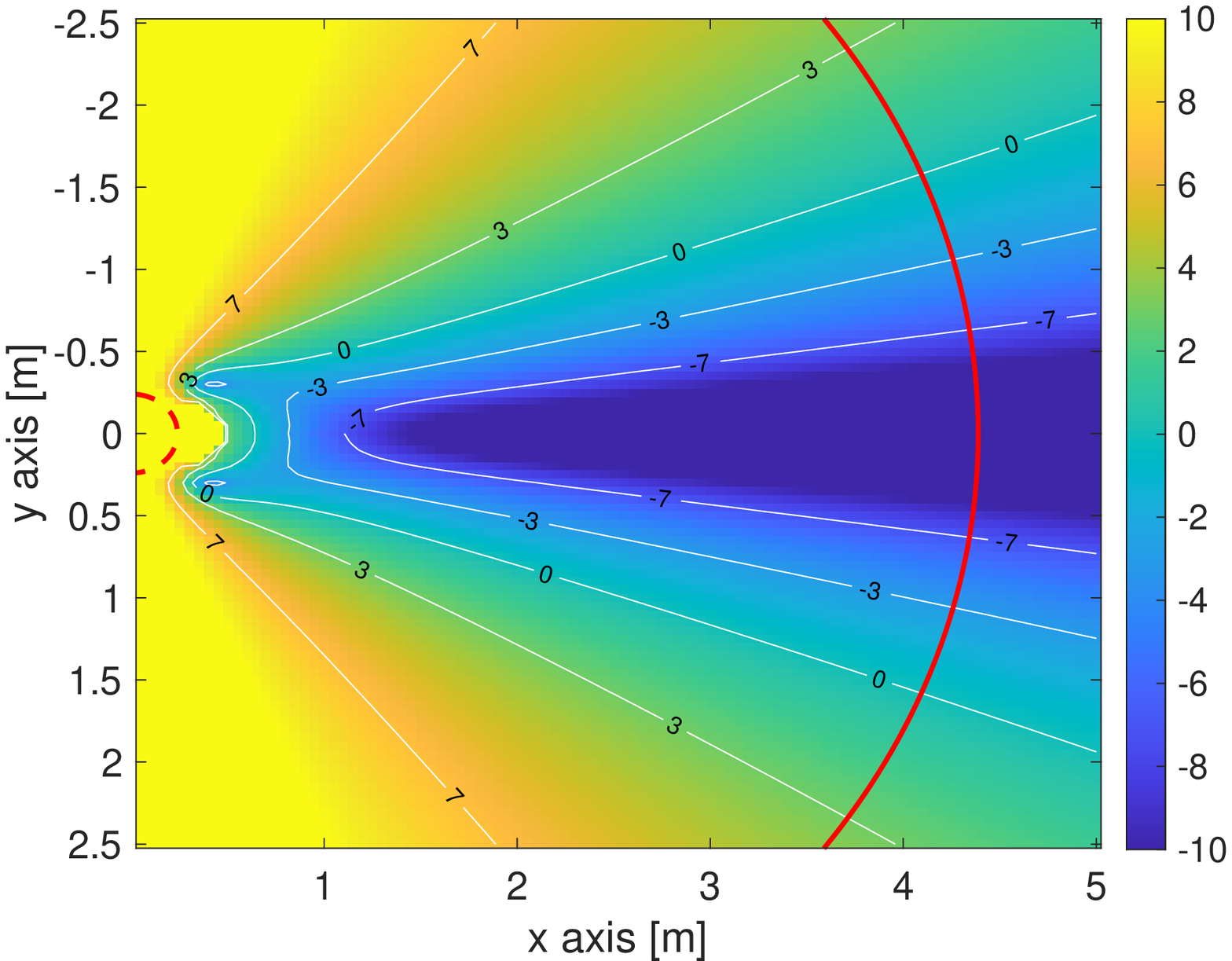}}
    % \include{Figures/fig-2-1}
    % \vspace{-1cm}
    \centerline{\small{(a) $P=\unit[30]{dBm}$}} \medskip
\end{minipage}
\hfill
\begin{minipage}[b]{0.245\linewidth}
  \centering
  \centerline{\includegraphics[width=1.1\linewidth]{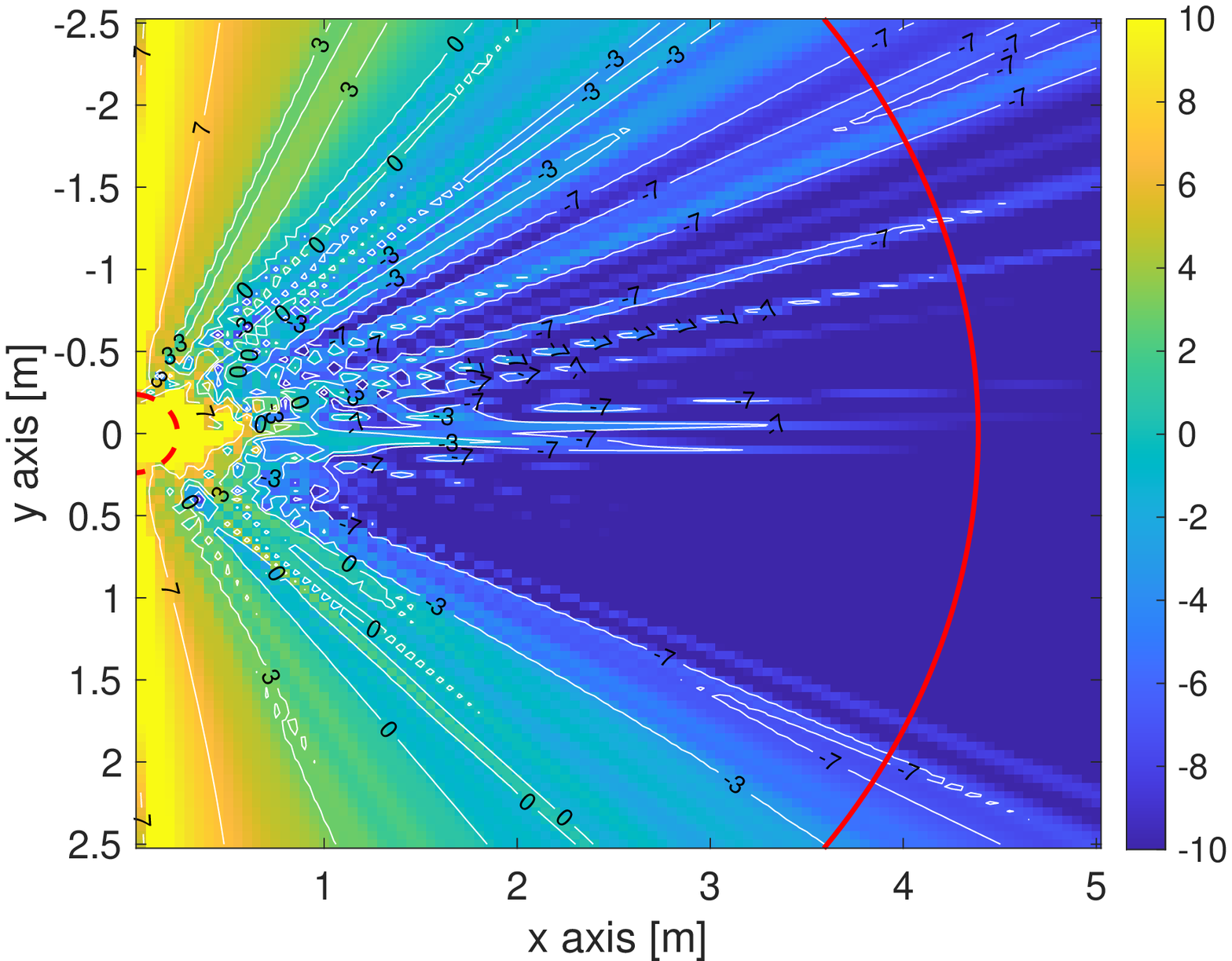}}
    % \include{Figures/fig-2-2}
    % \vspace{-1cm}
    \centerline{\small{(b) Analog, $G=50$}}\medskip
\end{minipage}
\begin{minipage}[b]{0.245\linewidth}
  \centering
  \centerline{\includegraphics[width=1.1\linewidth]{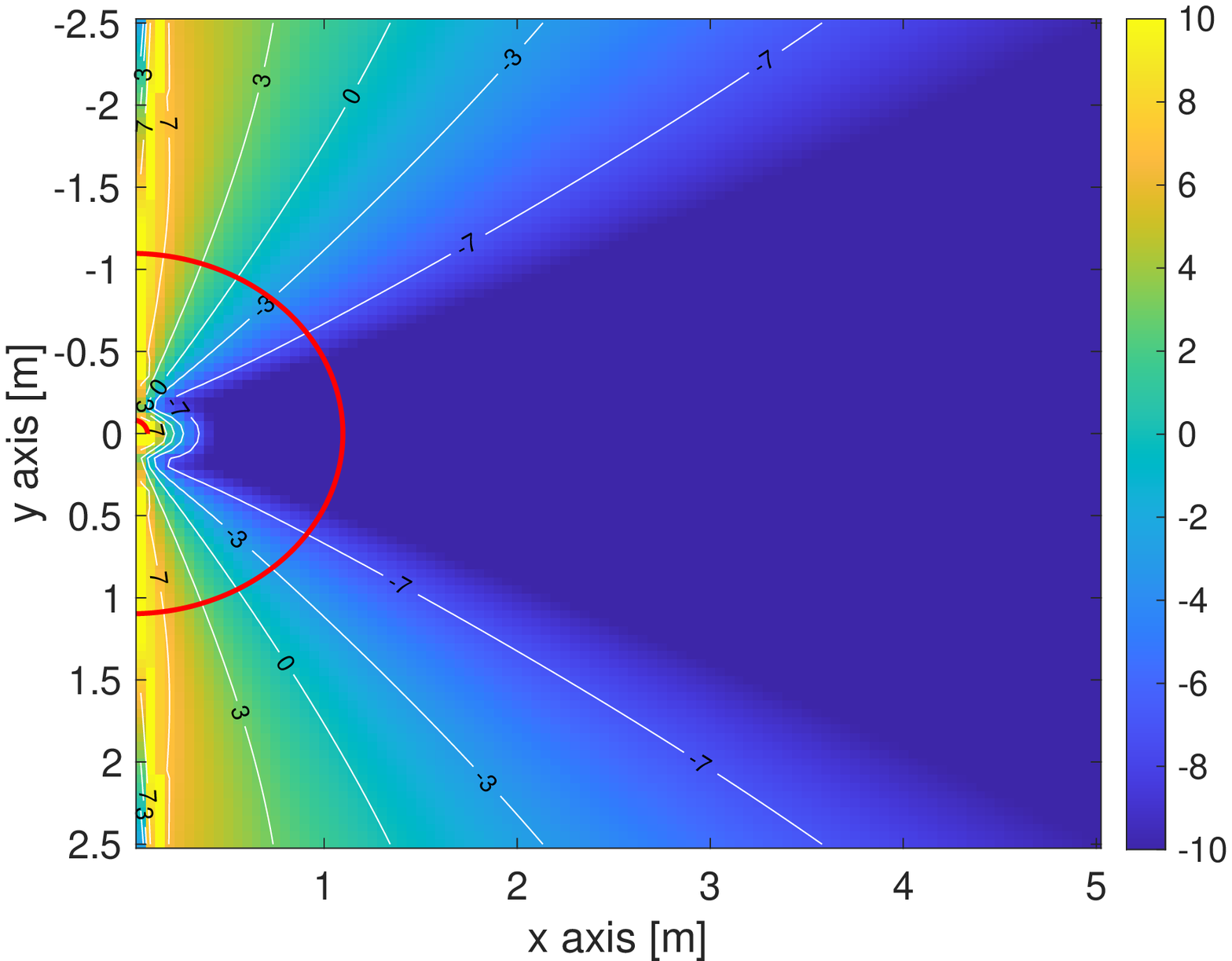}}
    % \include{Figures/fig-2-1}
    % \vspace{-1cm}
    \centerline{\small{(c) $N=32$}} \medskip
\end{minipage}
\hfill
\begin{minipage}[b]{0.245\linewidth}
  \centering
  \centerline{\includegraphics[width=1.1\linewidth]{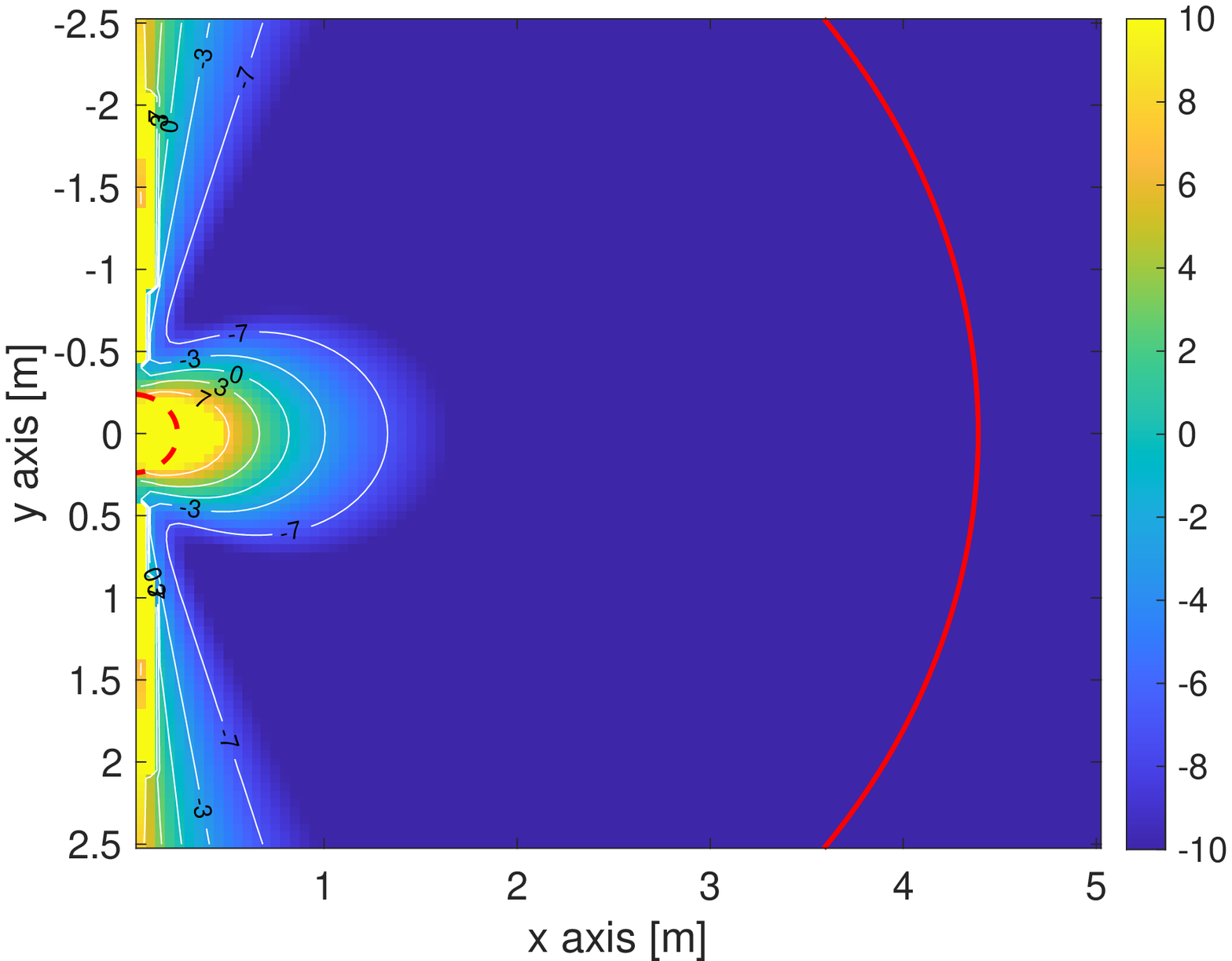}}
    % \include{Figures/fig-2-2}
    % \vspace{-1cm}
    \centerline{\small{(d) $W=\unit[100]{MHz}$}}\medskip
\end{minipage}
\vspace{-0.8cm}
\caption{Visualization of relative mismatch error MME-PEB (benchmarked by the MME-PEB in Fig.~\ref{fig:micro_scale} (a)) for different scenarios: (a) the transmission power $P$ is changed from $\unit[10]{dBm}$ to $\unit[30]{dBm}$; (b) the array structure is changed to analog array (single RFC) with $\mathcal{G} = 50$ transmissions; (c) the array size is changed from 64 to 32; (d) the bandwidth $W$ is changed from $\unit[400]{MHz}$ to $\unit[100]{MHz}$.}
\label{fig:different_scenarios}
\end{figure*}

% \begin{figure}[ht]
% \begin{minipage}[b]{0.78\linewidth}
%     \centering
%      \centerline{\includegraphics[width=0.98\linewidth]{Figures/n-1-4.png}}
% \end{minipage}
% \begin{minipage}[b]{0.78\linewidth}
%     \centering
%      \centerline{\includegraphics[width=0.98\linewidth]{Figures/n-2-4.png}}
% \end{minipage}
% % \vspace{-1cm}
% \caption{Comparison between simulation results and the derived lower bounds (LB, CRB-FF, and CRB-NF). The proposed localization algorithm attaches the bound when the average transmission power $P$ is greater than $\unit[2.5]{dBm}$, and the LB diverges from the CRBs when $P>\unit[10]{dBm}$.}
% % \label{fig:estimator_bounds_comparison}
% \end{figure}

\subsection{Evaluation of Model Mismatch Error}
We use the proposed MME to visualize the error caused by using a MM.
From Fig.~\ref{fig:micro_scale} (a), we can see that both angle and distance affect the MME. By further decomposing the PEB into a AEB and a DEB, which are shown in Fig.~\ref{fig:micro_scale} (b) and Fig.~\ref{fig:micro_scale} (c), it is observed that compared with angle estimation, the delay estimation is less affected by the model mismatch and the $\unit[-3]{dB}$ boundary is within $\unit[1]{m}$. However, the area MME-AEB $>\unit[-3]{dB}$ is much larger than that of the MME-PEB, indicating that the PEB is mainly determined by DEB in the NF.
% the angle error in the NF has limited contribution to the position estimation.

\subsection{MME for Different System Parameters}
In Fig.~\ref{fig:different_scenarios}, we compare the MME for different scenarios benchmarked by the MME-PEB in Fig.~\ref{fig:micro_scale} (a). We found out that the mismatch is getting larger with a higher transmission power $P$, as shown in (a). When an analog array is used {(and assumes a frequency-independent combiner matrix $\Wm_g$)}, the contours are not smooth due to randomness of the combiner matrix, but a similar mismatch pattern compared with a digital array can be seen with sufficient transmissions, as shown in (b). In Fig.~\ref{fig:different_scenarios} (c), the array size is changed from $64$ to $32$, and the area with model mismatch is largely reduced, indicating the effect of SNS and SWM are mitigated. When the bandwidth $W$ is changed from $\unit[400]{MHz}$ to $\unit[100]{MHz}$, the mismatch is reduced as the MM does not consider the BSE.

\section{Conclusion}
\label{sec:conclusion}
In this work, we derived the LB of a mismatched estimator using MCRB and analyzed the effect of different types of model impairments, namely, SNS, SWM, and BSE. A \ac{mme} model mismatch error is further defined as the absolute difference between the LB and the CRB of the TM, normalized by the CRB of the TM, which is determined by system parameters. {From the analysis, we see that the SNS has the least contribution among all the model impairments, the SWM dominates when the distance is small compared to the array size, and BSE has a more significant effect when the distance is much larger than the array size.}
The analysis in this work can provide suggestions on the tradeoff between the complexity of channel model and performance loss (by using an approximated, simple model).
In future work, we would like to analyze the effect of model mismatch on the 3D position and 3D orientation of an unsynchronized system.

\section*{Acknowledgment}
This work was supported, in part, by the European Commission through the H2020 project Hexa-X (Grant Agreement no. 101015956) and by the MSCA-IF grant 888913 (OTFS-RADCOM), and by Academy of Finland Profi-5 (n:o 326346) and ULTRA (n:o 328215) projects.

\appendices
\section{}
\label{sec:appendix_A}
% \subsection{CRB for the TM}
To derive the CRB for the TM, we notice that only channel gain $\alpha_{k,n}$ in~\eqref{eq:near_field_channel_amplitude} is dependent of channel parameters $\rho$ and $\xi$ as $\frac{\partial \alpha_{k,n}}{\partial \rho} = \alpha_{k,n}/\rho$ and $\frac{\partial \alpha_{k,n}}{\partial \xi} = -j\alpha_{k,n}$.
The derivative of channel vector with respect to the states $\rho$, $\xi$ and $\pv$ can then be expressed as 
% $\frac{\partial h_{k,n}}{\partial \rho} = \frac{h_{k,n}}{\rho}$, 
% $\frac{\partial h_{k,n}}{\partial \xi} = -j{h_{k,n}}$, and 
% $\frac{\partial h_{k,n}}{\partial\pv} 
% = \frac{h_{k,n}}{\alpha_{k,n}(\pv)} \frac{\partial\alpha_{k,n}(\pv)}{\partial\pv} 
% + \frac{h_{k,n}}{d_{k,n}(\pv)} \frac{\partial d_{k,n}(\pv)}{\partial\pv} 
% + \frac{h_{k,n}}{D_{k}(\pv)} \frac{\partial D_{k}(\pv)}{\partial\pv}$.
% Based on these derivatives and~\eqref{eq:FIM}, the FIM of the state parameters can be obtained.
\footnotesize
\begin{align}
    \frac{\partial h_{k,n}}{\partial \rho} & = \frac{h_{k,n}}{\rho},\ \ \ 
    \frac{\partial h_{k,n}}{\partial \xi} = -j{h_{k,n}},
    \\
    \frac{\partial h_{k,n}}{\partial\pv} &
    = \frac{h_{k,n}}{\alpha_{k,n}(\pv)} \frac{\partial\alpha_{k,n}(\pv)}{\partial\pv} 
    + \frac{h_{k,n}}{d_{k,n}(\pv)} \frac{\partial d_{k,n}(\pv)}{\partial\pv} \\
    & + \frac{h_{k,n}}{D_{k}(\pv)} \frac{\partial D_{k}(\pv)}{\partial\pv}, \notag
\end{align}
\normalsize
where
\footnotesize
\begin{align}
    \frac{\partial \alpha_{k,n}(\pv)}{\partial \pv} & = \alpha\frac{\lambda_k}{\lambda_c} \left(\frac{\pv}{\Vert \pv\Vert \Vert\pv-\bv_n \Vert} - \frac{(\pv-\bv_n)\Vert \pv\Vert }{\Vert\pv-\bv_n \Vert^3}\right),
    \\
    \frac{\partial d_{k,n}(\pv)}{\partial \pv} & = -j\frac{2\pi}{\lambda_k}d_{k,n}(\pv)
    \left(\frac{\pv-\bv_n}{\Vert\pv-\bv_n \Vert} - \frac{\pv}{\Vert\pv\Vert}\right),
    \\
    \frac{\partial D_{k}(\pv)}{\partial \pv} & = -j\frac{2\pi}{\lambda_k}D_{k}(\pv)
    \frac{\pv}{\Vert\pv\Vert}.
\end{align}
\normalsize
Based on these derivatives and~\eqref{eq:FIM}, the FIM of the state parameters can be obtained.

\ifCLASSOPTIONcaptionsoff
\newpage
\fi
\balance
% \end{thebibliography}
\bibliographystyle{IEEEtran}
\bibliography{ref}

\end{document}

%% file: Acronyms.tex
\DeclareAcronym{5g}{
short=5G,
long= fifth generation,
}
\DeclareAcronym{6g}{
short=6G,
long= sixth generation,
}
\DeclareAcronym{2d}{
short=2D,
long= two-dimensional,
}
\DeclareAcronym{3d}{
short=3D,
long= three-dimensional,
}
\DeclareAcronym{aod}{
short=AOD,
long= angle-of-departure,
}
\DeclareAcronym{aosa}{
short=AOSA,
long= array-of-subarray,
}
\DeclareAcronym{adod}{
short=ADOD,
long= angle-difference-of-departure,
}
\DeclareAcronym{aoa}{
short=AOA,
long= angle-of-arrival,
}
\DeclareAcronym{adc}{
short=ADC,
long= analog to digital converter,
}
\DeclareAcronym{aeb}{
short=AEB,
long= angle error bound,
}
\DeclareAcronym{av}{
short=AV,
long= autonomous vehicle,
}
\DeclareAcronym{bs}{
short=BS,
long= base station,
}
\DeclareAcronym{bse}{
short=BSE,
long= beam squint effect,
}
\DeclareAcronym{csi}{
short=CSI,
long= channel state information,
}
\DeclareAcronym{cfo}{
short=CFO,
long= carrier frequency offset,
}
\DeclareAcronym{ceb}{
short=CEB,
long= clock error bound,
}
\DeclareAcronym{coa}{
short=COA,
long= curvature-of-arrival,
}
\DeclareAcronym{crb}{
short=CRB,
long= Cram\'er-Rao bound,
}
\DeclareAcronym{ccrb}{
short=CCRB,
long= constrained Cram\'er-Rao bound,
}
\DeclareAcronym{cmos}{
short=CMOS,
long= complementary metal-oxide-semiconductor,
}
\DeclareAcronym{crlb}{
short=CRLB,
long= Cram\'er-Rao lower bound,
}
\DeclareAcronym{cdf}{
short=CDF,
long= cumulative distribution function,
}
\DeclareAcronym{cp}{
short=CP,
long= cyclic prefix,
}
\DeclareAcronym{dac}{
short=DAC,
long= digital to analog converter,
}
\DeclareAcronym{dfl}{
short=DFL,
long= device-free localization,
}
\DeclareAcronym{dmimo}{
short=D-MIMO,
long= distributed MIMO,
}
\DeclareAcronym{dlprs}{
short=DL-PRS,
long= downlink positioning reference signal,
}
\DeclareAcronym{d2d}{
short=D2D,
long= device-to-device,
}
\DeclareAcronym{dftsofdm}{
short=DFT-s-OFDM,
long= discrete-Fourier-transform spread OFDM,
}
\DeclareAcronym{dl}{
short=DL,
long= deep learning,
}
\DeclareAcronym{gps}{
short=GPS,
long= global positioning system,
}
\DeclareAcronym{ff}{
short=FF,
long= far-field,
}
\DeclareAcronym{fim}{
short=FIM,
long= Fisher information matrix,
}
\DeclareAcronym{nf}{
short=NF,
long= near-field,
}
\DeclareAcronym{hwi}{
short=HWI,
long= hardware impairment,
}
\DeclareAcronym{hemt}{
short=HEMT,
long= high electron mobility transistor,
}
\DeclareAcronym{hbt}{
short=HBT,
long= heterojunction bipolar transistors,
}
\DeclareAcronym{iot}{
short=IoT,
long= internet of things,
}
\DeclareAcronym{isac}{
short=ISAC,
long= integrated sensing and communication,
}
\DeclareAcronym{iqi}{
short=IQI,
long= in-phase and quadrature imbalance,
}
\DeclareAcronym{ia}{
short=IA,
long= initial access,
}
\DeclareAcronym{kpi}{
short=KPI,
long= key performance indicator,
}
\DeclareAcronym{kf}{
short=KF,
long= Kalman filter,
}
\DeclareAcronym{ekf}{
short=EKF,
long= extended Kalman filter,
}
\DeclareAcronym{ukf}{
short=UKF,
long= unscented Kalman filter,
}
\DeclareAcronym{ckf}{
short=CKF,
long= cubature Kalman filter,
}
\DeclareAcronym{pf}{
short=PF,
long= particle filter,
}
\DeclareAcronym{lb}{
short=LB,
long= lower bound,
}
\DeclareAcronym{lse}{
short=LSE,
long= least-square estimator,
}
\DeclareAcronym{lo}{
short=LO,
long= local oscillator,
}
\DeclareAcronym{mc}{
short=MC,
long= mutual coupling,
}
\DeclareAcronym{mac}{
short=MAC,
long= medium access control,
}
\DeclareAcronym{meb}{
short=MEB,
long= mapping error bound,
}
\DeclareAcronym{ml}{
short=ML,
long= machine learning,
}
\DeclareAcronym{mcrb}{
short=MCRB,
long= misspecified Cram\'er-Rao bound,
}
\DeclareAcronym{mds}{
short=MDS,
long= multidimensional scaling ,
}
\DeclareAcronym{mimo}{
short=MIMO,
long= multiple-input-multiple-output,
}
\DeclareAcronym{mm}{
short=MM,
long= mismatched model,
}
\DeclareAcronym{mpc}{
short=MPC,
long= multipath components,
}
\DeclareAcronym{mmwave}{
short=mmWave,
long= millimeter wave,
}
\DeclareAcronym{mmle}{
short=MMLE,
long= mismatched maximum likelihood estimation,
}
\DeclareAcronym{mme}{
short=MME,
long= model-mismatch error,
}
\DeclareAcronym{mems}{
short=MEMS,
long= micro-electro-mechanical system,
}
\DeclareAcronym{mle}{
short=MLE,
long= maximum likelihood estimation,
}
\DeclareAcronym{nlos}{
short=NLOS,
long= non-line-of-sight,
}
\DeclareAcronym{ofdm}{
short=OFDM,
long= orthogonal frequency-division multiplexing,
}
\DeclareAcronym{oeb}{
short=OEB,
long= orientation error bound,
}
\DeclareAcronym{otfs}{
short=OTFS,
long= orthogonal time-frequency space,
}
\DeclareAcronym{pdf}{
short=PDF,
long= probability density function,
}
\DeclareAcronym{papr}{
short=PAPR,
long= peak-to-average-power ratio,
}
\DeclareAcronym{pan}{
short=PAN,
long= power amplifier nonlinearity,
}
\DeclareAcronym{pa}{
short=PA,
long= power amplifier,
}
\DeclareAcronym{ps}{
short=PS,
long= phase shifter,
}
\DeclareAcronym{pn}{
short=PN,
long= phase noise,
}
\DeclareAcronym{poa}{
short=POA,
long= phase-of-arrival,
}
\DeclareAcronym{pwm}{
short=PWM,
long= planar wave model,
}
\DeclareAcronym{pdoa}{
short=PDOA,
long= phase-difference-of-arrival,
}
\DeclareAcronym{prs}{
short=PRS,
long= positioning reference signals,
}
\DeclareAcronym{peb}{
short=PEB,
long= position error bound,
}
\DeclareAcronym{rnn}{
short=RNN,
long= recurrent neural network,
}
\DeclareAcronym{rl}{
short=RL,
long= reinforcement learning,
}
\DeclareAcronym{rfc}{
short=RFC,
long= radio-frequency chain,
}
\DeclareAcronym{rf}{
short=RF,
long= radio frequency,
}
\DeclareAcronym{rfid}{
short=RFID,
long= radio frequency identification,
}
\DeclareAcronym{ris}{
short=RIS,
long= reconfigurable intelligent surface,
}
\DeclareAcronym{rss}{
short=RSS,
long= received signal strength,
}
\DeclareAcronym{rtt}{
short=RTT,
long= round-trip time,
}
\DeclareAcronym{sm}{
short=SM,
long= standard model,
}
\DeclareAcronym{snr}{
short=SNR,
long= signal-to-noise ratio,
}
\DeclareAcronym{sige}{
short=SiGe,
long= silicon-germanium,
}
\DeclareAcronym{spp}{
short=SPP,
long= surface plasmon polariton,
}
\DeclareAcronym{sa}{
short=SA,
long= subarray,
}
\DeclareAcronym{sns}{
short=SNS,
long= spatial non-stationarity,
}
\DeclareAcronym{sota}{
short=SOTA,
long= state-of-the-art,
}
\DeclareAcronym{swm}{
short=SWM,
long= spherical wave model,
}
\DeclareAcronym{slam}{
short=SLAM,
long= simultaneous localization and mapping,
}
\DeclareAcronym{tm}{
short=TM,
long= true model,
}
\DeclareAcronym{toa}{
short=TOA,
long= time-of-arrival,
}
\DeclareAcronym{tof}{
short=TOF,
long= time-of-flight,
}
\DeclareAcronym{tdoa}{
short=TDOA,
long= time-difference-of-arrival,
}
\DeclareAcronym{thz}{
short=THz,
long= terahertz,
}
\DeclareAcronym{ue}{
short=UE,
long= user equipment,
}
\DeclareAcronym{ummimo}{
short=UM-MIMO,
long= ultra-massive multi-input-multi-output,
}
\DeclareAcronym{vlp}{
short=VLP,
long= visible light positioning,
}
\DeclareAcronym{veb}{
short=VEB,
long= velocity error bound,
}
\DeclareAcronym{vlc}{
short=VLC,
long= visible light communication,
}
\DeclareAcronym{ula}{
short=ULA,
long= uniform linear array,
}
\DeclareAcronym{upa}{
short=UPA,
long= uniform planar array,
}
\DeclareAcronym{wlan}{
short=WLAN,
long= wireless local area network,
}
\DeclareAcronym{xlmimo}{
short=XL-MIMO,
long= extra-large MIMO,
}

%% file: macros.tex
\long\def\comment#1{}

\newfont{\bbb}{msbm10 scaled 700}

\newfont{\bb}{msbm10 scaled 1100}

\newcommand{\av}{{\bf a}}
\newcommand{\bv}{{\bf b}}

\newcommand{\dv}{{\bf d}}

\newcommand{\hv}{{\bf h}}

\newcommand{\nv}{{\bf n}}

\newcommand{\pv}{{\bf p}}

\newcommand{\yv}{{\bf y}}

\newcommand{\Am}{{\bf A}}
\newcommand{\Bm}{{\bf B}}

\newcommand{\Jm}{{\bf J}}

\newcommand{\Wm}{{\bf W}}

\newcommand{\Gc}{{\cal G}}

\newcommand{\alphav}{\hbox{\boldmath$\alpha$}}

\newcommand{\etav}{\hbox{\boldmath$\eta$}}

\newcommand{\epsilonv}{\hbox{\boldmath$\epsilon$}}

\newcommand{\muv}{\hbox{\boldmath$\mu$}}

\newcommand{\thetav}{\hbox{$\boldsymbol\theta$}}

\newcommand{\trace}{{\hbox{tr}}}

\renewcommand{\arg}{{\hbox{arg}}}

%% file: Figures_tikz/Fig-1.tex
\begin{tikzpicture}[scale=1\columnwidth/10cm,font=\footnotesize]
\begin{axis}[%
width=8cm,
height=3.0cm,
scale only axis,
xmin=-10,
xmax=30,
xlabel style={font=\color{white!15!black}},
xlabel={$P$ [dBm]},
ymode=log,
ymin=0.000630957344480193,
ymax=1,
yminorticks=true,
ylabel style={font=\color{white!15!black}},
ylabel={Position RMSE [m]},
axis background/.style={fill=white},
xmajorgrids,
ymajorgrids,
% yminorgrids,
legend columns=2, 
legend style={font=\footnotesize, at={(0.49,0.55)}, anchor=south west, legend cell align=left, align=left, draw=white!15!black}
]
\addplot [color=blue, dashed, line width=1.0pt, mark size=3.0pt, mark=o, mark options={solid, blue}]
  table[row sep=crcr]{%
-10	0.431361656255769\\
-7.14285714285717	0.208517078772163\\
-4.28571428571428	0.0408007338581275\\
-1.42857142857144	0.0317719638127366\\
1.42857142857144	0.0213541410433545\\
4.28571428571428	0.0160345316898419\\
7.14285714285717	0.011569157770034\\
10	0.00905187722855507\\
12.8571428571428	0.00686259626578563\\
15.7142857142857	0.00531322615970575\\
18.5714285714286	0.00463860022629419\\
21.4285714285714	0.00419370065366498\\
24.2857142857143	0.00384784308220146\\
27.1428571428572	0.00372159328825493\\
30	0.00363610743350145\\
};
\addlegendentry{MMLE}

\addplot [color=red, dashed, line width=1.0pt, mark size=2.5pt, mark=square, mark options={solid, red}]
  table[row sep=crcr]{%
-10	0.424617726390332\\
-7.14285714285717	0.195991960835085\\
-4.28571428571428	0.0402825286533734\\
-1.42857142857144	0.0312468688959112\\
1.42857142857144	0.0209111802335275\\
4.28571428571428	0.0156170150355109\\
7.14285714285717	0.0109052527199325\\
10	0.00832616841148139\\
12.8571428571428	0.00581306655332172\\
15.7142857142857	0.00397639310784893\\
18.5714285714286	0.00298914571530319\\
21.4285714285714	0.00220187312384928\\
24.2857142857143	0.00149682613247258\\
27.1428571428572	0.00110527278444911\\
30	0.000769691086616372\\
};
\addlegendentry{MLE-TM}

\addplot [color=black, line width=1.0pt, mark size=3.0pt, mark=+, mark options={solid, black}]
  table[row sep=crcr]{%
-10	0.0796227025349349\\
-7.14285714285717	0.05735632413492\\
-4.28571428571428	0.0413520765692297\\
-1.42857142857144	0.0298624315729062\\
1.42857142857144	0.0216324978585438\\
4.28571428571428	0.0157625791365163\\
7.14285714285717	0.0116088884952034\\
10	0.00871088358647143\\
12.8571428571428	0.00673641034390937\\
15.7142857142857	0.00543890499087342\\
18.5714285714286	0.00462594782482117\\
21.4285714285714	0.00414263137625249\\
24.2857142857143	0.00386863466425702\\
27.1428571428572	0.00371879120273288\\
30	0.00363875528235874\\
};
\addlegendentry{LB}

\addplot [color=blue, line width=1.0pt, mark size=2.7pt, mark=triangle, mark options={solid, blue}]
  table[row sep=crcr]{%
-10	0.0792473495314325\\
-7.14285714285717	0.0570331820850308\\
-4.28571428571428	0.0410459640319139\\
-1.42857142857144	0.0295401922564134\\
1.42857142857144	0.0212596531277917\\
4.28571428571428	0.0153002677791209\\
7.14285714285717	0.0110113835080929\\
10	0.0079247349541267\\
12.8571428571428	0.00570331820730883\\
15.7142857142857	0.00410459640380034\\
18.5714285714286	0.00295401922447706\\
21.4285714285714	0.00212596531384682\\
24.2857142857143	0.00153002677746563\\
27.1428571428572	0.00110113835145094\\
30	0.000792473495459377\\
};
\addlegendentry{CRB-MM}

\addplot [color=red, line width=1.0pt]
  table[row sep=crcr]{%
-10	0.079338486074432\\
-7.14285714285717	0.0570987717396979\\
-4.28571428571428	0.0410931679684402\\
-1.42857142857144	0.0295741642443407\\
1.42857142857144	0.0212841022997644\\
4.28571428571428	0.0153178634863029\\
7.14285714285717	0.0110240468946822\\
10	0.00793384860767582\\
12.8571428571428	0.00570987717520978\\
15.7142857142857	0.00410931679794486\\
18.5714285714286	0.00295741642565504\\
21.4285714285714	0.00212841023071007\\
24.2857142857143	0.0015317863489129\\
27.1428571428572	0.00110240468965333\\
30	0.00079338486100306\\
};
\addlegendentry{CRB-TM}
\end{axis}

\end{tikzpicture}%

%% file: Figures_tikz/Fig-5-1-a.tex
\begin{tikzpicture}
[scale=1\columnwidth/10cm,font=\normalsize]
\begin{axis}[%
width=8cm,
height=4cm,
at={(0, 0)},
scale only axis,
xmin=0,
xmax=150,
xlabel style={font=\large\color{white!15!black}, yshift=1 ex},xlabel={UE Speed [m/s]},
xlabel={Array Size},
ymin=-80,
ymax=0,
ylabel style={font=\large\color{white!15!black}, yshift= -1.5 ex},
ylabel={Mismatch Error [dB]},
axis background/.style={fill=white},
xmajorgrids,
ymajorgrids,
legend style={font=\Large, at={(0.03, 1.2)}, anchor=south west, legend cell align=left, align=left, draw=white!15!black, legend columns=4}
]
\addplot [color=black, line width=1.5pt]
  table[row sep=crcr]{%
4	-30.3389733560306\\
9	-26.513125845497\\
16	-23.8714338540216\\
25	-21.8942168294953\\
36	-20.2616562078701\\
49	-18.7520093422323\\
64	-17.1627559521306\\
81	-15.3122481650645\\
100	-13.1113723936606\\
121	-10.5951343927858\\
144	-7.85410875955159\\
};
\addlegendentry{MME (TM) $\ \ $}

\addplot [color=blue, dashed, line width=1.5pt, mark=o, mark options={solid, blue}]
  table[row sep=crcr]{%
4	-71.1943012263596\\
9	-66.0495246567927\\
16	-61.0779843034366\\
25	-57.2285550628704\\
36	-54.0916880999675\\
49	-51.4264782717509\\
64	-49.1150847638211\\
81	-47.0721875921469\\
100	-45.2452475156404\\
121	-43.5887647175142\\
144	-42.0776240121915\\
};
\addlegendentry{MME (TM-SNS) $\ \ $}

\addplot [color=red, dashed, line width=1.5pt, mark=square, mark options={solid, red}]
  table[row sep=crcr]{%
4	-73.9412947218889\\
9	-56.9773138259053\\
16	-46.711974496114\\
25	-38.8920415588318\\
36	-32.5296203470013\\
49	-27.1520522674617\\
64	-22.4873366221104\\
81	-18.3588791835254\\
100	-14.6402996567144\\
121	-11.2312360597974\\
144	-8.0411488320326\\
};
\addlegendentry{MME (TM-SWM) $\ \ $}

\addplot [color=green, dashed, line width=1.5pt, mark=diamond, mark options={solid, green}]
  table[row sep=crcr]{%
4	-30.3403598084575\\
9	-26.5045618402122\\
16	-23.8441900110111\\
25	-21.8528620023309\\
36	-20.2507897466915\\
49	-18.9061350012181\\
64	-17.7461066514874\\
81	-16.7257005219467\\
100	-15.8148846778654\\
121	-14.9925178468029\\
144	-14.2431153369961\\
};
\addlegendentry{MME (TM-BSE) $\ \ $}

\end{axis}

% \begin{axis}[%
% width=7.778in,
% height=5.833in,
% at={(0in,0in)},
% scale only axis,
% xmin=0,
% xmax=1,
% ymin=0,
% ymax=1,
% axis line style={draw=none},
% ticks=none,
% axis x line*=bottom,
% axis y line*=left
% ]
% \end{axis}
\end{tikzpicture}%

%% file: Figures_tikz/Fig-5-2-a.tex
\begin{tikzpicture}
[scale=1\columnwidth/10cm,font=\normalsize]
\begin{axis}[%
width=8cm,
height=4cm,
at={(0, 0)},
scale only axis,
xmode = log,
xmin=0.25,
xmax=10,
xlabel style={font=\large\color{white!15!black}, yshift=1 ex},
xlabel={Distance},
ymin=-60,
ymax=20,
ylabel style={font=\large\color{white!15!black}, yshift= -1.5 ex},
ylabel={Mismatch Error [dB]},
axis background/.style={fill=white},
xmajorgrids,
ymajorgrids,
]
\addplot [color=black, line width=1.5pt, forget plot]
  table[row sep=crcr]{%
0.1	29.4518235256342\\
0.104761575278966	37.5803601260501\\
0.109749876549306	34.6736492266037\\
0.114975699539774	34.2636999472553\\
0.120450354025878	33.8040788370022\\
0.126185688306602	32.4746232215748\\
0.132194114846603	32.1421683256039\\
0.138488637139387	28.1409978300135\\
0.145082877849594	27.713078371085\\
0.151991108295293	27.3546616482686\\
0.159228279334109	27.0181896587115\\
0.166810053720006	26.699064907556\\
0.174752840000768	26.3722482404346\\
0.183073828029537	26.0397240539376\\
0.191791026167249	25.8187551326865\\
0.200923300256505	25.4657547105253\\
0.210490414451202	25.1022242391119\\
0.220513073990305	25.3316028630389\\
0.231012970008316	24.8824444991859\\
0.242012826479438	12.5465824598048\\
0.253536449397011	11.8551694252888\\
0.265608778294669	11.2073233341897\\
0.278255940220712	10.5409155125454\\
0.291505306282518	9.8534523615603\\
0.305385550883342	9.15066344374458\\
0.319926713779738	8.43790596566407\\
0.335160265093884	7.71966699844427\\
0.351119173421513	6.99894064244364\\
0.367837977182863	6.27680779468909\\
0.385352859371053	5.55682204240793\\
0.403701725859655	4.83751689366454\\
0.42292428743895	4.12019253098447\\
0.443062145758388	3.40460326902422\\
0.464158883361278	2.6899949435778\\
0.486260158006535	1.97735691064287\\
0.509413801481638	1.26542800451034\\
0.533669923120631	0.553591883712727\\
0.559081018251222	-0.158805355327843\\
0.585702081805667	-0.872445028237848\\
0.613590727341317	-1.58628478877068\\
0.642807311728432	-2.3011100657281\\
0.673415065775082	-3.0176936791168\\
0.705480231071864	-3.73305653227754\\
0.739072203352578	-4.44910849075649\\
0.774263682681127	-5.16307251750436\\
0.811130830789687	-5.87385656828849\\
0.849753435908644	-6.58013138013212\\
0.890215085445039	-7.27947213729664\\
0.93260334688322	-7.97011228071835\\
0.977009957299225	-8.64977046531459\\
1.02353102189903	-9.31587515695196\\
1.07226722201032	-9.96649281876529\\
1.12332403297803	-10.5973976693277\\
1.176811952435	-11.207195132272\\
1.23284673944207	-11.7930128958993\\
1.29154966501488	-12.3521926886071\\
1.35304777457981	-12.8820159467047\\
1.41747416292681	-13.3808815277779\\
1.48496826225447	-13.8472561253276\\
1.55567614393047	-14.2796873214367\\
1.62975083462064	-14.6777481056987\\
1.70735264747069	-15.0412248116521\\
1.78864952905743	-15.3707586031787\\
1.87381742286038	-15.6691648405853\\
1.96304065004027	-15.9325102100185\\
2.05651230834865	-16.1700381317357\\
2.15443469003188	-16.3760040185215\\
2.25701971963392	-16.5586826886337\\
2.36448941264541	-16.718425571411\\
2.47707635599171	-16.8594730886687\\
2.59502421139974	-16.9803033498314\\
2.71858824273294	-17.084956153774\\
2.8480358684358	-17.175194966739\\
2.98364724028334	-17.2532073476465\\
3.12571584968824	-17.316548764788\\
3.27454916287773	-17.3785141069956\\
3.43046928631492	-17.4284916513348\\
3.59381366380463	-17.471680739643\\
3.76493580679247	-17.5076531869621\\
3.94420605943766	-17.5414541547562\\
4.13201240011534	-17.569625552326\\
4.32876128108306	-17.5941902055032\\
4.53487850812858	-17.6127192432967\\
4.7508101621028	-17.6347532044398\\
4.97702356433211	-17.6516008024346\\
5.21400828799968	-17.6665600546492\\
5.46227721768434	-17.6775695336909\\
5.72236765935022	-17.6924466434894\\
5.99484250318941	-17.7037686974364\\
6.28029144183425	-17.7120012534733\\
6.57933224657568	-17.7220455502533\\
6.8926121043497	-17.7314685921774\\
7.22080901838546	-17.74192425717\\
7.56463327554629	-17.7507231444575\\
7.92482898353917	-17.7594263760475\\
8.30217568131975	-17.7681542411813\\
8.69749002617784	-17.7769871162398\\
9.11162756115489	-17.7860506481821\\
9.54548456661834	-17.7954274389082\\
10	-17.8052232874807\\
};
\addplot [color=blue, dashed, line width=1.5pt, forget plot]
  table[row sep=crcr]{%
0.1	-20.2511542259351\\
0.104761575278966	-20.6441160179525\\
0.109749876549306	-21.0374788726521\\
0.114975699539774	-21.4313105666366\\
0.120450354025878	-21.8256504802923\\
0.126185688306602	-22.2205443820038\\
0.132194114846603	-22.6159855226712\\
0.138488637139387	-23.0119907923702\\
0.145082877849594	-23.408526819691\\
0.151991108295293	-23.8055897891318\\
0.159228279334109	-24.2031708449225\\
0.166810053720006	-24.6012075702264\\
0.174752840000768	-24.9997273435093\\
0.183073828029537	-25.3986611685154\\
0.191791026167249	-25.798005700387\\
0.200923300256505	-26.1976989452894\\
0.210490414451202	-26.5977904002584\\
0.220513073990305	-26.9981660028321\\
0.231012970008316	-27.3988529461205\\
0.242012826479438	-27.799823739515\\
0.253536449397011	-28.2010656616157\\
0.265608778294669	-28.6025426415135\\
0.278255940220712	-29.004201384453\\
0.291505306282518	-29.406105395577\\
0.305385550883342	-29.8081600679657\\
0.319926713779738	-30.210361397465\\
0.335160265093884	-30.6128394451766\\
0.351119173421513	-31.0153002358709\\
0.367837977182863	-31.417954136527\\
0.385352859371053	-31.820690586921\\
0.403701725859655	-32.2235880370525\\
0.42292428743895	-32.6265719577088\\
0.443062145758388	-33.0296021970555\\
0.464158883361278	-33.4328378332611\\
0.486260158006535	-33.836016490987\\
0.509413801481638	-34.2393302420998\\
0.533669923120631	-34.642690744904\\
0.559081018251222	-35.0461759202018\\
0.585702081805667	-35.4495556551017\\
0.613590727341317	-35.8531986724763\\
0.642807311728432	-36.2565924194155\\
0.673415065775082	-36.6604781722374\\
0.705480231071864	-37.0642221854461\\
0.739072203352578	-37.4675088392964\\
0.774263682681127	-37.8713486107481\\
0.811130830789687	-38.2746662574027\\
0.849753435908644	-38.6784081895212\\
0.890215085445039	-39.0827831851757\\
0.93260334688322	-39.4865522345962\\
0.977009957299225	-39.8902170815842\\
1.02353102189903	-40.2945960079484\\
1.07226722201032	-40.6977154032624\\
1.12332403297803	-41.1014967783746\\
1.176811952435	-41.5060627364591\\
1.23284673944207	-41.9093398858421\\
1.29154966501488	-42.3123209845488\\
1.35304777457981	-42.7167883888997\\
1.41747416292681	-43.1203680993404\\
1.48496826225447	-43.5259331869484\\
1.55567614393047	-43.9278944821734\\
1.62975083462064	-44.3303549044788\\
1.70735264747069	-44.7348324377762\\
1.78864952905743	-45.1396829928816\\
1.87381742286038	-45.544332506281\\
1.96304065004027	-45.9491187708229\\
2.05651230834865	-46.3522004619203\\
2.15443469003188	-46.7548132831841\\
2.25701971963392	-47.1597508644886\\
2.36448941264541	-47.5634849625368\\
2.47707635599171	-47.9680596020625\\
2.59502421139974	-48.3742581456404\\
2.71858824273294	-48.7708906039774\\
2.8480358684358	-49.177071133844\\
2.98364724028334	-49.5860438982359\\
3.12571584968824	-49.9859193454452\\
3.27454916287773	-50.3858567202922\\
3.43046928631492	-50.7897048617589\\
3.59381366380463	-51.1977161304441\\
3.76493580679247	-51.6067678613322\\
3.94420605943766	-51.9992229359336\\
4.13201240011534	-52.3890077203297\\
4.32876128108306	-52.8023836043893\\
4.53487850812858	-53.2106806261046\\
4.7508101621028	-53.6100654812484\\
4.97702356433211	-54.0168727010396\\
5.21400828799968	-54.4091380568558\\
5.46227721768434	-54.8165471707722\\
5.72236765935022	-55.2294802238438\\
5.99484250318941	-55.6367720197274\\
6.28029144183425	-56.011017934636\\
6.57933224657568	-56.4275622080967\\
6.8926121043497	-56.848604485778\\
7.22080901838546	-57.2500166214081\\
7.56463327554629	-57.6458509199907\\
7.92482898353917	-58.0401800624661\\
8.30217568131975	-58.4080582365405\\
8.69749002617784	-58.8675731969922\\
9.11162756115489	-59.2605953358689\\
9.54548456661834	-59.6695344502243\\
10	-60.0286176912894\\
};
\addplot [color=red, dashed, line width=1.5pt, forget plot]
  table[row sep=crcr]{%
0.1	28.3160069052796\\
0.104761575278966	27.7713502850531\\
0.109749876549306	27.1976675341866\\
0.114975699539774	32.9239176191618\\
0.120450354025878	32.4871575661894\\
0.126185688306602	30.6726615842863\\
0.132194114846603	30.3002856840833\\
0.138488637139387	27.1091051098182\\
0.145082877849594	26.4282582640121\\
0.151991108295293	25.9585786326306\\
0.159228279334109	25.5495207965379\\
0.166810053720006	25.1668300616177\\
0.174752840000768	24.7950988157141\\
0.183073828029537	24.4275008724373\\
0.191791026167249	24.0602107681584\\
0.200923300256505	23.6900252277807\\
0.210490414451202	23.3026119636484\\
0.220513073990305	22.925502973145\\
0.231012970008316	22.5398635652809\\
0.242012826479438	11.0784558732344\\
0.253536449397011	10.3694003431623\\
0.265608778294669	9.84937302592328\\
0.278255940220712	9.32461993878599\\
0.291505306282518	8.76929512653987\\
0.305385550883342	8.18683848729711\\
0.319926713779738	7.58417324961027\\
0.335160265093884	6.96709244941002\\
0.351119173421513	6.33978247845766\\
0.367837977182863	5.70508909615097\\
0.385352859371053	5.06484715311944\\
0.403701725859655	4.42014680542326\\
0.42292428743895	3.77155195915668\\
0.443062145758388	3.1192457552271\\
0.464158883361278	2.46316128279096\\
0.486260158006535	1.80308130488515\\
0.509413801481638	1.13871962059699\\
0.533669923120631	0.469811319501332\\
0.559081018251222	-0.203851654290304\\
0.585702081805667	-0.882362349460948\\
0.613590727341317	-1.56566565464107\\
0.642807311728432	-2.25353888880295\\
0.673415065775082	-2.94559043086679\\
0.705480231071864	-3.64126457720027\\
0.739072203352578	-4.33986179720955\\
0.774263682681127	-5.04055388648595\\
0.811130830789687	-5.7424436461319\\
0.849753435908644	-6.44455214530643\\
0.890215085445039	-7.14590445126692\\
0.93260334688322	-7.84549790639909\\
0.977009957299225	-8.54238545818129\\
1.02353102189903	-9.23564456392796\\
1.07226722201032	-9.92442340853705\\
1.12332403297803	-10.6079116418194\\
1.176811952435	-11.2854053964819\\
1.23284673944207	-11.9562493405564\\
1.29154966501488	-12.6198627759561\\
1.35304777457981	-13.2757547594395\\
1.41747416292681	-13.9235004186328\\
1.48496826225447	-14.5627367936367\\
1.55567614393047	-15.1931754527437\\
1.62975083462064	-15.8145965372474\\
1.70735264747069	-16.4268218179554\\
1.78864952905743	-17.0297773236364\\
1.87381742286038	-17.6234252423397\\
1.96304065004027	-18.207751264944\\
2.05651230834865	-18.7828201363576\\
2.15443469003188	-19.3487605703206\\
2.25701971963392	-19.9057141675303\\
2.36448941264541	-20.4538815459751\\
2.47707635599171	-20.9934347146413\\
2.59502421139974	-21.5247067433615\\
2.71858824273294	-22.0479542295739\\
2.8480358684358	-22.56338366976\\
2.98364724028334	-23.0714606167597\\
3.12571584968824	-23.5723281778002\\
3.27454916287773	-24.0664998532299\\
3.43046928631492	-24.5541835429696\\
3.59381366380463	-25.0357542751312\\
3.76493580679247	-25.5116508420745\\
3.94420605943766	-25.981944933531\\
4.13201240011534	-26.4472758976303\\
4.32876128108306	-26.9079884086465\\
4.53487850812858	-27.3636466750023\\
4.7508101621028	-27.8153674685196\\
4.97702356433211	-28.2632623791298\\
5.21400828799968	-28.7073362677282\\
5.46227721768434	-29.1482791874142\\
5.72236765935022	-29.5857252497866\\
5.99484250318941	-30.0198913557125\\
6.28029144183425	-30.4515818272703\\
6.57933224657568	-30.8805717404547\\
6.8926121043497	-31.3072549611891\\
7.22080901838546	-31.7305179626546\\
7.56463327554629	-32.1537916191588\\
7.92482898353917	-32.5746634037936\\
8.30217568131975	-32.993197634391\\
8.69749002617784	-33.4106583896291\\
9.11162756115489	-33.8262299585523\\
9.54548456661834	-34.2371433797932\\
10	-34.6506592665688\\
};
\addplot [color=green, dashed, line width=1.5pt, forget plot]
  table[row sep=crcr]{%
0.1	-17.7368186023428\\
0.104761575278966	-17.7378422511478\\
0.109749876549306	-17.7386900681188\\
0.114975699539774	-17.7388182331558\\
0.120450354025878	-17.738492899478\\
0.126185688306602	-17.7376670360175\\
0.132194114846603	-17.7385897612467\\
0.138488637139387	-17.7381091255598\\
0.145082877849594	-17.7383130858099\\
0.151991108295293	-17.7389376171512\\
0.159228279334109	-17.7387937333405\\
0.166810053720006	-17.7388036157717\\
0.174752840000768	-17.7384754398228\\
0.183073828029537	-17.7384965046912\\
0.191791026167249	-17.7387457771466\\
0.200923300256505	-17.7383622901526\\
0.210490414451202	-17.738791478321\\
0.220513073990305	-17.7388251617612\\
0.231012970008316	-17.7388445149918\\
0.242012826479438	-17.7389558100692\\
0.253536449397011	-17.7387400314959\\
0.265608778294669	-17.7389519148934\\
0.278255940220712	-17.738864733917\\
0.291505306282518	-17.7389332872557\\
0.305385550883342	-17.7390690468658\\
0.319926713779738	-17.7390720369638\\
0.335160265093884	-17.7390711436895\\
0.351119173421513	-17.7390962640353\\
0.367837977182863	-17.7390426195914\\
0.385352859371053	-17.7391174513208\\
0.403701725859655	-17.7390363677737\\
0.42292428743895	-17.739093394945\\
0.443062145758388	-17.7390870636955\\
0.464158883361278	-17.739206858147\\
0.486260158006535	-17.7392329405876\\
0.509413801481638	-17.7392041377829\\
0.533669923120631	-17.739224716717\\
0.559081018251222	-17.7392256552955\\
0.585702081805667	-17.7393018505614\\
0.613590727341317	-17.739349880098\\
0.642807311728432	-17.7393592136216\\
0.673415065775082	-17.7393988901196\\
0.705480231071864	-17.7394849033134\\
0.739072203352578	-17.7395289193955\\
0.774263682681127	-17.7395254832707\\
0.811130830789687	-17.7396163055545\\
0.849753435908644	-17.7396881162391\\
0.890215085445039	-17.7397311913744\\
0.93260334688322	-17.7397590752572\\
0.977009957299225	-17.7398175755092\\
1.02353102189903	-17.7399198231783\\
1.07226722201032	-17.7400626601593\\
1.12332403297803	-17.7401506997384\\
1.176811952435	-17.7402703802862\\
1.23284673944207	-17.740373256538\\
1.29154966501488	-17.7405091878067\\
1.35304777457981	-17.7406621267322\\
1.41747416292681	-17.7407796620647\\
1.48496826225447	-17.7409562310245\\
1.55567614393047	-17.7411816958362\\
1.62975083462064	-17.7413894454997\\
1.70735264747069	-17.7415887006304\\
1.78864952905743	-17.7418346273936\\
1.87381742286038	-17.7421573712944\\
1.96304065004027	-17.7424049696194\\
2.05651230834865	-17.74277307945\\
2.15443469003188	-17.7431400998852\\
2.25701971963392	-17.7435249155099\\
2.36448941264541	-17.744000661653\\
2.47707635599171	-17.7444569366367\\
2.59502421139974	-17.7449887936258\\
2.71858824273294	-17.7455646519444\\
2.8480358684358	-17.7462069952737\\
2.98364724028334	-17.7469079330702\\
3.12571584968824	-17.7476648947801\\
3.27454916287773	-17.7485120590678\\
3.43046928631492	-17.7494258442258\\
3.59381366380463	-17.750470433671\\
3.76493580679247	-17.7515548582916\\
3.94420605943766	-17.7527880007293\\
4.13201240011534	-17.754115718423\\
4.32876128108306	-17.7555917956778\\
4.53487850812858	-17.7572244825605\\
4.7508101621028	-17.7590008426157\\
4.97702356433211	-17.7609197177151\\
5.21400828799968	-17.7630464211276\\
5.46227721768434	-17.765382707782\\
5.72236765935022	-17.7679383594847\\
5.99484250318941	-17.7707503639229\\
6.28029144183425	-17.7738322506968\\
6.57933224657568	-17.7772059545702\\
6.8926121043497	-17.7809116803722\\
7.22080901838546	-17.7849686343803\\
7.56463327554629	-17.7894232899439\\
7.92482898353917	-17.794306909331\\
8.30217568131975	-17.79965599283\\
8.69749002617784	-17.8055051625579\\
9.11162756115489	-17.8119391884939\\
9.54548456661834	-17.8189821579653\\
10	-17.8266942306279\\
};
\addplot [color=blue, only marks, line width=1.5pt, mark=o, mark options={solid, blue}, forget plot]
  table[row sep=crcr]{%
0.1	-20.2511542259351\\
0.120450354025878	-21.8256504802923\\
0.145082877849594	-23.408526819691\\
0.174752840000768	-24.9997273435093\\
0.210490414451202	-26.5977904002584\\
0.253536449397011	-28.2010656616157\\
0.305385550883342	-29.8081600679657\\
0.367837977182863	-31.417954136527\\
0.443062145758388	-33.0296021970555\\
0.533669923120631	-34.642690744904\\
0.642807311728432	-36.2565924194155\\
0.774263682681127	-37.8713486107481\\
0.93260334688322	-39.4865522345962\\
1.12332403297803	-41.1014967783746\\
1.35304777457981	-42.7167883888997\\
1.62975083462064	-44.3303549044788\\
1.96304065004027	-45.9491187708229\\
2.36448941264541	-47.5634849625368\\
2.8480358684358	-49.177071133844\\
3.43046928631492	-50.7897048617589\\
4.13201240011534	-52.3890077203297\\
4.97702356433211	-54.0168727010396\\
5.99484250318941	-55.6367720197274\\
7.22080901838546	-57.2500166214081\\
8.69749002617784	-58.8675731969922\\
};
\addplot [color=red, only marks, line width=1.5pt, mark=square, mark options={solid, red}, forget plot]
  table[row sep=crcr]{%
0.1	28.3160069052796\\
0.120450354025878	32.4871575661894\\
0.145082877849594	26.4282582640121\\
0.174752840000768	24.7950988157141\\
0.210490414451202	23.3026119636484\\
0.253536449397011	10.3694003431623\\
0.305385550883342	8.18683848729711\\
0.367837977182863	5.70508909615097\\
0.443062145758388	3.1192457552271\\
0.533669923120631	0.469811319501332\\
0.642807311728432	-2.25353888880295\\
0.774263682681127	-5.04055388648595\\
0.93260334688322	-7.84549790639909\\
1.12332403297803	-10.6079116418194\\
1.35304777457981	-13.2757547594395\\
1.62975083462064	-15.8145965372474\\
1.96304065004027	-18.207751264944\\
2.36448941264541	-20.4538815459751\\
2.8480358684358	-22.56338366976\\
3.43046928631492	-24.5541835429696\\
4.13201240011534	-26.4472758976303\\
4.97702356433211	-28.2632623791298\\
5.99484250318941	-30.0198913557125\\
7.22080901838546	-31.7305179626546\\
8.69749002617784	-33.4106583896291\\
};
\addplot [color=green, only marks, line width=1.5pt, mark=diamond, mark options={solid, green}, forget plot]
  table[row sep=crcr]{%
0.1	-17.7368186023428\\
0.120450354025878	-17.738492899478\\
0.145082877849594	-17.7383130858099\\
0.174752840000768	-17.7384754398228\\
0.210490414451202	-17.738791478321\\
0.253536449397011	-17.7387400314959\\
0.305385550883342	-17.7390690468658\\
0.367837977182863	-17.7390426195914\\
0.443062145758388	-17.7390870636955\\
0.533669923120631	-17.739224716717\\
0.642807311728432	-17.7393592136216\\
0.774263682681127	-17.7395254832707\\
0.93260334688322	-17.7397590752572\\
1.12332403297803	-17.7401506997384\\
1.35304777457981	-17.7406621267322\\
1.62975083462064	-17.7413894454997\\
1.96304065004027	-17.7424049696194\\
2.36448941264541	-17.744000661653\\
2.8480358684358	-17.7462069952737\\
3.43046928631492	-17.7494258442258\\
4.13201240011534	-17.754115718423\\
4.97702356433211	-17.7609197177151\\
5.99484250318941	-17.7707503639229\\
7.22080901838546	-17.7849686343803\\
8.69749002617784	-17.8055051625579\\
};
\end{axis}

\end{tikzpicture}%

%% file: Figures_tikz/Fig-5-1-b.tex
\begin{tikzpicture}
[scale=1\columnwidth/10cm,font=\normalsize]
\begin{axis}[%
width=8cm,
height=4cm,
at={(0, 0)},
scale only axis,
xmin=0,
xmax=150,
xlabel style={font=\large\color{white!15!black}, yshift=1 ex},xlabel={UE Speed [m/s]},
xlabel={Array Size},
ymin=-100,
ymax=20,
ylabel style={font=\large\color{white!15!black}, yshift= -1.5 ex},
ylabel={Mismatch Error [dB]},
axis background/.style={fill=white},
xmajorgrids,
ymajorgrids,
]
\addplot [color=black, line width=1.5pt,]
  table[row sep=crcr]{%
4	-25.3859384170382\\
9	-15.6488533558333\\
16	-8.47588533117749\\
25	-3.34873389553286\\
36	0.297605705333297\\
49	2.95347834546868\\
64	4.9456050712551\\
81	6.448071310047\\
100	7.52806888943536\\
121	8.16386133287864\\
144	8.24467272639635\\
};
% \addlegendentry{LB (TM)}

\addplot [color=blue,  dashed,  line width=1.5pt, mark=o, mark options={solid, blue}]
  table[row sep=crcr]{%
4	-69.6605015872358\\
9	-65.7942488001075\\
16	-61.6638444651392\\
25	-58.0568813942048\\
36	-54.9907694824853\\
49	-52.3571158345367\\
64	-50.0591503052823\\
81	-48.0247987933652\\
100	-46.2013969217352\\
121	-44.550099435921\\
144	-43.0416177068628\\
};
% \addlegendentry{LB (only SNS)}

\addplot [color=red, dashed,  line width=1.5pt, mark=square, mark options={solid, red}]
  table[row sep=crcr]{%
4	-85.5694095496214\\
9	-57.7018535390999\\
16	-46.2949428128825\\
25	-36.865231132024\\
36	-28.3838122323638\\
49	-20.3389605761135\\
64	-12.9165048255615\\
81	-6.41527389674036\\
100	-1.15145754139019\\
121	2.88597491996393\\
144	6.00022943503387\\
};
% \addlegendentry{LB (only SWM)}

\addplot [color=green, dashed,  line width=1.5pt, mark=diamond, mark options={solid, green}]
  table[row sep=crcr]{%
4	-25.3870644024763\\
9	-15.6367789672208\\
16	-8.42963130312398\\
25	-3.2412829798713\\
36	0.499864680544874\\
49	3.29583309710313\\
64	5.4999909613183\\
81	7.32081544130176\\
100	8.87819395301926\\
121	10.2440212418814\\
144	11.4639609764591\\
};
% \addlegendentry{LB (only BSE)}

\end{axis}

% \begin{axis}[%
% width=7.778in,
% height=5.833in,
% at={(0in,0in)},
% scale only axis,
% xmin=0,
% xmax=1,
% ymin=0,
% ymax=1,
% axis line style={draw=none},
% ticks=none,
% axis x line*=bottom,
% axis y line*=left
% ]
% \end{axis}
\end{tikzpicture}%

%% file: Figures_tikz/Fig-5-2-b.tex
\begin{tikzpicture}
[scale=1\columnwidth/10cm,font=\normalsize]
\begin{axis}[%
width=8cm,
height=4cm,
at={(0, 0)},
scale only axis,
xmode = log,
xmin=0.25,
xmax=10,
xlabel style={font=\large\color{white!15!black}, yshift=1 ex},
xlabel={Distance},
ymin=-60,
ymax=30,
ylabel style={font=\large\color{white!15!black}, yshift= -1.5 ex},
ylabel={Mismatch Error [dB]},
axis background/.style={fill=white},
xmajorgrids,
ymajorgrids,
]
\addplot [color=black, line width=1.5pt, forget plot]
  table[row sep=crcr]{%
0.1	39.2693054978631\\
0.104761575278966	47.4354663370092\\
0.109749876549306	44.7348312474439\\
0.114975699539774	44.5640814435303\\
0.120450354025878	44.3695684713584\\
0.126185688306602	42.7477932388275\\
0.132194114846603	42.6868640435084\\
0.138488637139387	38.8800236216674\\
0.145082877849594	38.8038836478789\\
0.151991108295293	38.7549957728785\\
0.159228279334109	38.7065110798453\\
0.166810053720006	38.6501609615405\\
0.174752840000768	38.5834587549226\\
0.183073828029537	38.5055353501194\\
0.191791026167249	38.4167423716538\\
0.200923300256505	38.317383248423\\
0.210490414451202	38.207731208878\\
0.220513073990305	39.0588536230308\\
0.231012970008316	38.8445336917571\\
0.242012826479438	26.4028769354915\\
0.253536449397011	26.2131300178878\\
0.265608778294669	25.8345889171627\\
0.278255940220712	25.3639088418645\\
0.291505306282518	24.8391225399706\\
0.305385550883342	24.2782917586624\\
0.319926713779738	23.6908705219016\\
0.335160265093884	23.0820742750574\\
0.351119173421513	22.454781324843\\
0.367837977182863	21.810441948303\\
0.385352859371053	21.1495430524735\\
0.403701725859655	20.4718317777058\\
0.42292428743895	19.776429832161\\
0.443062145758388	19.0618494425335\\
0.464158883361278	18.3259498374844\\
0.486260158006535	17.5658367266465\\
0.509413801481638	16.7776640080507\\
0.533669923120631	15.9563476632832\\
0.559081018251222	15.0951042310143\\
0.585702081805667	14.1863864493931\\
0.613590727341317	13.2141598989482\\
0.642807311728432	12.1593381683985\\
0.673415065775082	10.9920496087523\\
0.705480231071864	9.6813474706538\\
0.739072203352578	8.13429287455949\\
0.774263682681127	6.18891536848093\\
0.811130830789687	3.45120675004296\\
0.849753435908644	-1.4918054809545\\
0.890215085445039	-11.8033108698569\\
0.93260334688322	-1.86848815468682\\
0.977009957299225	1.68405915553961\\
1.02353102189903	3.44167178742708\\
1.07226722201032	4.49008889498992\\
1.12332403297803	5.20392432340361\\
1.176811952435	5.68600789993167\\
1.23284673944207	6.02218923293304\\
1.29154966501488	6.25452295055698\\
1.35304777457981	6.409442231235\\
1.41747416292681	6.50451126880489\\
1.48496826225447	6.55204764425503\\
1.55567614393047	6.56102775851366\\
1.62975083462064	6.53811732991664\\
1.70735264747069	6.48845310498791\\
1.78864952905743	6.41601957136947\\
1.87381742286038	6.32125084120139\\
1.96304065004027	6.2149317126644\\
2.05651230834865	6.08852059572303\\
2.15443469003188	5.95367815847305\\
2.25701971963392	5.80463372523184\\
2.36448941264541	5.64497580236298\\
2.47707635599171	5.47385680113598\\
2.59502421139974	5.29590589642396\\
2.71858824273294	5.10987983648823\\
2.8480358684358	4.91650573881407\\
2.98364724028334	4.71613659557192\\
3.12571584968824	4.51255883307238\\
3.27454916287773	4.29631553047625\\
3.43046928631492	4.07762657740307\\
3.59381366380463	3.85340182779953\\
3.76493580679247	3.62496621408474\\
3.94420605943766	3.38929897831963\\
4.13201240011534	3.14979855171822\\
4.32876128108306	2.9055454222617\\
4.53487850812858	2.65892053886074\\
4.7508101621028	2.40311154422244\\
4.97702356433211	2.14512811874727\\
5.21400828799968	1.88280920492156\\
5.46227721768434	1.61797967111949\\
5.72236765935022	1.34492095009676\\
5.99484250318941	1.06952035195013\\
6.28029144183425	0.791617626802065\\
6.57933224657568	0.50757668220507\\
6.8926121043497	0.219435923288358\\
7.22080901838546	-0.0739901606016815\\
7.56463327554629	-0.37046808716417\\
7.92482898353917	-0.671102272013343\\
8.30217568131975	-0.975880915950654\\
8.69749002617784	-1.28475765563854\\
9.11162756115489	-1.59770986307336\\
9.54548456661834	-1.91469253287579\\
10	-2.23566273711589\\
};
\addplot [color=blue, dashed, line width=1.5pt, forget plot]
  table[row sep=crcr]{%
0.1	-21.3633523222206\\
0.104761575278966	-21.7396973029455\\
0.109749876549306	-22.1186574933759\\
0.114975699539774	-22.4999779052246\\
0.120450354025878	-22.8834266236894\\
0.126185688306602	-23.2688106216715\\
0.132194114846603	-23.6559112351341\\
0.138488637139387	-24.0445976900623\\
0.145082877849594	-24.4346947621781\\
0.151991108295293	-24.8260794407281\\
0.159228279334109	-25.2186284582744\\
0.166810053720006	-25.6122311886177\\
0.174752840000768	-26.0067857926446\\
0.183073828029537	-26.4022063404114\\
0.191791026167249	-26.7984082006476\\
0.200923300256505	-27.1953217115573\\
0.210490414451202	-27.5928782487392\\
0.220513073990305	-27.9910201762174\\
0.231012970008316	-28.3897007578652\\
0.242012826479438	-28.7888443453958\\
0.253536449397011	-29.1884329881984\\
0.265608778294669	-29.5884186579904\\
0.278255940220712	-29.9887754908736\\
0.291505306282518	-30.3894395137406\\
0.305385550883342	-30.7904039408279\\
0.319926713779738	-31.1916407918571\\
0.335160265093884	-31.5931233216039\\
0.351119173421513	-31.9948235591031\\
0.367837977182863	-32.3967256372734\\
0.385352859371053	-32.7988172235425\\
0.403701725859655	-33.2010600697241\\
0.42292428743895	-33.6034553877191\\
0.443062145758388	-34.0059873285022\\
0.464158883361278	-34.4086523247618\\
0.486260158006535	-34.811411923592\\
0.509413801481638	-35.2142686652312\\
0.533669923120631	-35.6172011294999\\
0.559081018251222	-36.0202169566026\\
0.585702081805667	-36.4232983215402\\
0.613590727341317	-36.826470696531\\
0.642807311728432	-37.229658447366\\
0.673415065775082	-37.6328932618527\\
0.705480231071864	-38.0361722666576\\
0.739072203352578	-38.4394869743128\\
0.774263682681127	-38.8428427237725\\
0.811130830789687	-39.246215797467\\
0.849753435908644	-39.6495727969306\\
0.890215085445039	-40.0529395118283\\
0.93260334688322	-40.45631409044\\
0.977009957299225	-40.8596939661946\\
1.02353102189903	-41.2630707757473\\
1.07226722201032	-41.6664247887744\\
1.12332403297803	-42.0697233242357\\
1.176811952435	-42.4730278874023\\
1.23284673944207	-42.8762981569141\\
1.29154966501488	-43.2795211973936\\
1.35304777457981	-43.6827170465637\\
1.41747416292681	-44.0858156578873\\
1.48496826225447	-44.4888858156826\\
1.55567614393047	-44.891837879608\\
1.62975083462064	-45.2947470452119\\
1.70735264747069	-45.6975854029024\\
1.78864952905743	-46.1002776071425\\
1.87381742286038	-46.5028954456736\\
1.96304065004027	-46.9053238811536\\
2.05651230834865	-47.3076943015632\\
2.15443469003188	-47.7098733007889\\
2.25701971963392	-48.1118511538336\\
2.36448941264541	-48.513658700372\\
2.47707635599171	-48.9152764007446\\
2.59502421139974	-49.3166596134435\\
2.71858824273294	-49.7178260235345\\
2.8480358684358	-50.1186651719218\\
2.98364724028334	-50.519273440467\\
3.12571584968824	-50.9195294877459\\
3.27454916287773	-51.3193495549522\\
3.43046928631492	-51.718848368586\\
3.59381366380463	-52.1178973971425\\
3.76493580679247	-52.5164127708122\\
3.94420605943766	-52.9144755869007\\
4.13201240011534	-53.3119315164942\\
4.32876128108306	-53.7088128704961\\
4.53487850812858	-54.104997267496\\
4.7508101621028	-54.500379470832\\
4.97702356433211	-54.8949325202769\\
5.21400828799968	-55.2886343281504\\
5.46227721768434	-55.6813361009228\\
5.72236765935022	-56.0730663870761\\
5.99484250318941	-56.4634713195508\\
6.28029144183425	-56.8527093996326\\
6.57933224657568	-57.2404985173465\\
6.8926121043497	-57.626841847075\\
7.22080901838546	-58.0114717897877\\
7.56463327554629	-58.3943602617984\\
7.92482898353917	-58.7754091760112\\
8.30217568131975	-59.1541372936724\\
8.69749002617784	-59.530499339769\\
9.11162756115489	-59.9048517661527\\
9.54548456661834	-60.2762299100453\\
10	-60.6446190442017\\
};
\addplot [color=red, dashed, line width=1.5pt, forget plot]
  table[row sep=crcr]{%
0.1	39.3367398485844\\
0.104761575278966	39.0636802957727\\
0.109749876549306	38.7526214878549\\
0.114975699539774	44.7993162051336\\
0.120450354025878	44.6240641983393\\
0.126185688306602	43.0430614954973\\
0.132194114846603	42.9282712895128\\
0.138488637139387	39.9227699474343\\
0.145082877849594	39.4916280524365\\
0.151991108295293	39.2764534828594\\
0.159228279334109	39.1196134842402\\
0.166810053720006	38.9835019740328\\
0.174752840000768	38.8535840860073\\
0.183073828029537	38.7232395837196\\
0.191791026167249	38.5890009461614\\
0.200923300256505	38.4488753508439\\
0.210490414451202	38.3014333328163\\
0.220513073990305	38.1451600092059\\
0.231012970008316	37.9777993024353\\
0.242012826479438	18.9812806589973\\
0.253536449397011	17.9168149234709\\
0.265608778294669	20.8881680546836\\
0.278255940220712	21.6769680368905\\
0.291505306282518	21.8572427397152\\
0.305385550883342	21.7636579310533\\
0.319926713779738	21.5197132945772\\
0.335160265093884	21.1831235641488\\
0.351119173421513	20.7848813726537\\
0.367837977182863	20.3432526629196\\
0.385352859371053	19.869738612693\\
0.403701725859655	19.371926338285\\
0.42292428743895	18.8550646279765\\
0.443062145758388	18.3228434757453\\
0.464158883361278	17.7779473753105\\
0.486260158006535	17.2223241613502\\
0.509413801481638	16.6573898074955\\
0.533669923120631	16.0842482528765\\
0.559081018251222	15.5036476442177\\
0.585702081805667	14.9161122749882\\
0.613590727341317	14.3219815960812\\
0.642807311728432	13.7214386235862\\
0.673415065775082	13.1145605320361\\
0.705480231071864	12.5012089042542\\
0.739072203352578	11.8811512312594\\
0.774263682681127	11.2541213820212\\
0.811130830789687	10.6195250941732\\
0.849753435908644	9.97693230299529\\
0.890215085445039	9.32542029924258\\
0.93260334688322	8.6642861703034\\
0.977009957299225	7.99232927963214\\
1.02353102189903	7.30839598827293\\
1.07226722201032	6.61102399770874\\
1.12332403297803	5.89885645481195\\
1.176811952435	5.16992832229806\\
1.23284673944207	4.4224098613314\\
1.29154966501488	3.65441569936741\\
1.35304777457981	2.86372275703042\\
1.41747416292681	2.04821831718436\\
1.48496826225447	1.20612943242611\\
1.55567614393047	0.335673652720288\\
1.62975083462064	-0.564091044334589\\
1.70735264747069	-1.49353031524805\\
1.78864952905743	-2.45277851458519\\
1.87381742286038	-3.44117999088945\\
1.96304065004027	-4.45552844485093\\
2.05651230834865	-5.49276046768261\\
2.15443469003188	-6.5506666153799\\
2.25701971963392	-7.62360888034215\\
2.36448941264541	-8.70952420917155\\
2.47707635599171	-9.80019004419711\\
2.59502421139974	-10.8950493525394\\
2.71858824273294	-11.9900797029717\\
2.8480358684358	-13.0753423619295\\
2.98364724028334	-14.1559653862896\\
3.12571584968824	-15.2168733059338\\
3.27454916287773	-16.2649886745452\\
3.43046928631492	-17.2908947074948\\
3.59381366380463	-18.2913723017937\\
3.76493580679247	-19.2724713631072\\
3.94420605943766	-20.2104739244945\\
4.13201240011534	-21.1296532809493\\
4.32876128108306	-22.0314601263662\\
4.53487850812858	-22.8512985788649\\
4.7508101621028	-23.6583030199897\\
4.97702356433211	-24.4394044989562\\
5.21400828799968	-25.1741631430735\\
5.46227721768434	-25.8927926020617\\
5.72236765935022	-26.5581666062795\\
5.99484250318941	-27.1715979077482\\
6.28029144183425	-27.7762106953798\\
6.57933224657568	-28.3536806854717\\
6.8926121043497	-28.9047726706917\\
7.22080901838546	-29.3796973454189\\
7.56463327554629	-29.9496801689391\\
7.92482898353917	-30.4743623846444\\
8.30217568131975	-30.9612319003301\\
8.69749002617784	-31.4569639412155\\
9.11162756115489	-31.9385039292166\\
9.54548456661834	-32.2997369879146\\
10	-32.7966405078482\\
};
\addplot [color=green, dashed, line width=1.5pt, forget plot]
  table[row sep=crcr]{%
0.1	20.9514001368311\\
0.104761575278966	20.7477304301683\\
0.109749876549306	20.5439928547144\\
0.114975699539774	20.3401840129347\\
0.120450354025878	20.1362677498035\\
0.126185688306602	19.9322734595469\\
0.132194114846603	19.7281506515814\\
0.138488637139387	19.5239424464594\\
0.145082877849594	19.3196628292936\\
0.151991108295293	19.1152416487996\\
0.159228279334109	18.9107053885132\\
0.166810053720006	18.7060632590849\\
0.174752840000768	18.5013273875141\\
0.183073828029537	18.2964146001665\\
0.191791026167249	18.0913859678814\\
0.200923300256505	17.8862138195237\\
0.210490414451202	17.6808706796775\\
0.220513073990305	17.4753849325115\\
0.231012970008316	17.2696894428069\\
0.242012826479438	17.0638678817274\\
0.253536449397011	16.8578632453935\\
0.265608778294669	16.6517005590628\\
0.278255940220712	16.4453056721518\\
0.291505306282518	16.2386683699401\\
0.305385550883342	16.0318449951072\\
0.319926713779738	15.8248030593577\\
0.335160265093884	15.6175015340904\\
0.351119173421513	15.4099788775523\\
0.367837977182863	15.2021905525173\\
0.385352859371053	14.9940725465239\\
0.403701725859655	14.7857486892912\\
0.42292428743895	14.5770800376634\\
0.443062145758388	14.3680853056604\\
0.464158883361278	14.1587460558605\\
0.486260158006535	13.9490776209979\\
0.509413801481638	13.7390451057495\\
0.533669923120631	13.5286755730441\\
0.559081018251222	13.3178727758655\\
0.585702081805667	13.1066353839199\\
0.613590727341317	12.894969980892\\
0.642807311728432	12.682861638648\\
0.673415065775082	12.470266113431\\
0.705480231071864	12.2571307461325\\
0.739072203352578	12.0435101875046\\
0.774263682681127	11.8293622964233\\
0.811130830789687	11.6145835794803\\
0.849753435908644	11.399227525105\\
0.890215085445039	11.1832484616503\\
0.93260334688322	10.9666217370999\\
0.977009957299225	10.749286008224\\
1.02353102189903	10.5312068313347\\
1.07226722201032	10.3123568951733\\
1.12332403297803	10.0927358793389\\
1.176811952435	9.87228459496981\\
1.23284673944207	9.65097310171234\\
1.29154966501488	9.42874109599131\\
1.35304777457981	9.20555238942887\\
1.41747416292681	8.98139419267886\\
1.48496826225447	8.75617611153785\\
1.55567614393047	8.5298477732656\\
1.62975083462064	8.3023982124526\\
1.70735264747069	8.07378023721435\\
1.78864952905743	7.84390392930851\\
1.87381742286038	7.61269599208214\\
1.96304065004027	7.38019247181455\\
2.05651230834865	7.14622125958454\\
2.15443469003188	6.91078648015508\\
2.25701971963392	6.6738195623077\\
2.36448941264541	6.43521212345382\\
2.47707635599171	6.19496398800169\\
2.59502421139974	5.95294831497824\\
2.71858824273294	5.70910837545682\\
2.8480358684358	5.46337002980573\\
2.98364724028334	5.21565354985245\\
3.12571584968824	4.96589196411367\\
3.27454916287773	4.71397972338883\\
3.43046928631492	4.45986408817731\\
3.59381366380463	4.20341375408248\\
3.76493580679247	3.94461781317528\\
3.94420605943766	3.68332033269314\\
4.13201240011534	3.41947928963793\\
4.32876128108306	3.15297744218918\\
4.53487850812858	2.88373426871334\\
4.7508101621028	2.61168407459405\\
4.97702356433211	2.33674547548031\\
5.21400828799968	2.05880903164245\\
5.46227721768434	1.77780248409442\\
5.72236765935022	1.49365594895664\\
5.99484250318941	1.20628914025298\\
6.28029144183425	0.915634655752399\\
6.57933224657568	0.621632984090979\\
6.8926121043497	0.324219952677311\\
7.22080901838546	0.0233510144495312\\
7.56463327554629	-0.281025168162159\\
7.92482898353917	-0.588941967469339\\
8.30217568131975	-0.900423193951269\\
8.69749002617784	-1.21547845722358\\
9.11162756115489	-1.53413827706862\\
9.54548456661834	-1.85638295809434\\
10	-2.18219529978751\\
};
\addplot [color=blue, only marks, line width=1.5pt, mark=o, mark options={solid, blue}, forget plot]
  table[row sep=crcr]{%
0.1	-21.3633523222206\\
0.120450354025878	-22.8834266236894\\
0.145082877849594	-24.4346947621781\\
0.174752840000768	-26.0067857926446\\
0.210490414451202	-27.5928782487392\\
0.253536449397011	-29.1884329881984\\
0.305385550883342	-30.7904039408279\\
0.367837977182863	-32.3967256372734\\
0.443062145758388	-34.0059873285022\\
0.533669923120631	-35.6172011294999\\
0.642807311728432	-37.229658447366\\
0.774263682681127	-38.8428427237725\\
0.93260334688322	-40.45631409044\\
1.12332403297803	-42.0697233242357\\
1.35304777457981	-43.6827170465637\\
1.62975083462064	-45.2947470452119\\
1.96304065004027	-46.9053238811536\\
2.36448941264541	-48.513658700372\\
2.8480358684358	-50.1186651719218\\
3.43046928631492	-51.718848368586\\
4.13201240011534	-53.3119315164942\\
4.97702356433211	-54.8949325202769\\
5.99484250318941	-56.4634713195508\\
7.22080901838546	-58.0114717897877\\
8.69749002617784	-59.530499339769\\
};
\addplot [color=red, only marks, line width=1.5pt, mark=square, mark options={solid, red}, forget plot]
  table[row sep=crcr]{%
0.1	39.3367398485844\\
0.120450354025878	44.6240641983393\\
0.145082877849594	39.4916280524365\\
0.174752840000768	38.8535840860073\\
0.210490414451202	38.3014333328163\\
0.253536449397011	17.9168149234709\\
0.305385550883342	21.7636579310533\\
0.367837977182863	20.3432526629196\\
0.443062145758388	18.3228434757453\\
0.533669923120631	16.0842482528765\\
0.642807311728432	13.7214386235862\\
0.774263682681127	11.2541213820212\\
0.93260334688322	8.6642861703034\\
1.12332403297803	5.89885645481195\\
1.35304777457981	2.86372275703042\\
1.62975083462064	-0.564091044334589\\
1.96304065004027	-4.45552844485093\\
2.36448941264541	-8.70952420917155\\
2.8480358684358	-13.0753423619295\\
3.43046928631492	-17.2908947074948\\
4.13201240011534	-21.1296532809493\\
4.97702356433211	-24.4394044989562\\
5.99484250318941	-27.1715979077482\\
7.22080901838546	-29.3796973454189\\
8.69749002617784	-31.4569639412155\\
};
\addplot [color=green, only marks, line width=1.5pt, mark=diamond, mark options={solid, green}, forget plot]
  table[row sep=crcr]{%
0.1	20.9514001368311\\
0.120450354025878	20.1362677498035\\
0.145082877849594	19.3196628292936\\
0.174752840000768	18.5013273875141\\
0.210490414451202	17.6808706796775\\
0.253536449397011	16.8578632453935\\
0.305385550883342	16.0318449951072\\
0.367837977182863	15.2021905525173\\
0.443062145758388	14.3680853056604\\
0.533669923120631	13.5286755730441\\
0.642807311728432	12.682861638648\\
0.774263682681127	11.8293622964233\\
0.93260334688322	10.9666217370999\\
1.12332403297803	10.0927358793389\\
1.35304777457981	9.20555238942887\\
1.62975083462064	8.3023982124526\\
1.96304065004027	7.38019247181455\\
2.36448941264541	6.43521212345382\\
2.8480358684358	5.46337002980573\\
3.43046928631492	4.45986408817731\\
4.13201240011534	3.41947928963793\\
4.97702356433211	2.33674547548031\\
5.99484250318941	1.20628914025298\\
7.22080901838546	0.0233510144495312\\
8.69749002617784	-1.21547845722358\\
};
\end{axis}

\end{tikzpicture}%

%% file: Figures_tikz/Fig-5-1-c.tex
\begin{tikzpicture}
[scale=1\columnwidth/10cm,font=\normalsize]
\begin{axis}[%
width=8cm,
height=4cm,
at={(0, 0)},
scale only axis,
xmin=0,
xmax=150,
xlabel style={font=\large\color{white!15!black}, yshift=1 ex},xlabel={UE Speed [m/s]},
xlabel={Array Size},
ymin=-80,
ymax=0,
ylabel style={font=\large\color{white!15!black}, yshift= -1.5 ex},
ylabel={Mismatch Error [dB]},
axis background/.style={fill=white},
xmajorgrids,
ymajorgrids,
]
\addplot [color=black, line width=1.5pt,]
  table[row sep=crcr]{%
4	-65.1640563417497\\
9	-54.8406442143007\\
16	-45.982550710161\\
25	-38.5974676744019\\
36	-32.4026837934291\\
49	-27.1070721164726\\
64	-22.4871878351468\\
81	-18.3862200633052\\
100	-14.6861263848583\\
121	-11.2885457803498\\
144	-8.10040025345796\\
};
% \addlegendentry{LB (TM)}

\addplot [color=blue, dashed, line width=1.5pt, mark=o, mark options={solid, blue}]
  table[row sep=crcr]{%
4	-72.1311169112416\\
9	-66.0205647840112\\
16	-61.0401907839645\\
25	-57.2255221042551\\
36	-54.0850628472338\\
49	-51.4243152212891\\
64	-49.1124382482031\\
81	-47.0717910564813\\
100	-45.2451285494185\\
121	-43.5887021146195\\
144	-42.0772143455547\\
};
% \addlegendentry{LB (only SNS)}

\addplot [color=red, dashed, line width=1.5pt, mark=square, mark options={solid, red}]
  table[row sep=crcr]{%
4	-72.686025244361\\
9	-56.9196952683141\\
16	-46.7245329790236\\
25	-38.9214110402841\\
36	-32.5675267952424\\
49	-27.200977690543\\
64	-22.5494849435819\\
81	-18.4355470656376\\
100	-14.7310842024752\\
121	-11.3328144368602\\
144	-8.14477028277639\\
};
% \addlegendentry{LB (only SWM)}

\addplot [color=green, dashed, line width=1.5pt, mark=diamond, mark options={solid, green}]
  table[row sep=crcr]{%
4	-66.4308745111663\\
9	-59.0919371429102\\
16	-54.075224053072\\
25	-50.1790160121292\\
36	-47.0074290664085\\
49	-44.3317592158506\\
64	-42.0110719016825\\
81	-39.9646921691664\\
100	-38.1338184385923\\
121	-36.4775146503138\\
144	-34.9655501645683\\
};
% \addlegendentry{LB (only BSE)}

\end{axis}

% \begin{axis}[%
% width=7.778in,
% height=5.833in,
% at={(0in,0in)},
% scale only axis,
% xmin=0,
% xmax=1,
% ymin=0,
% ymax=1,
% axis line style={draw=none},
% ticks=none,
% axis x line*=bottom,
% axis y line*=left
% ]
% \end{axis}
\end{tikzpicture}%

%% file: Figures_tikz/Fig-5-2-c.tex
\begin{tikzpicture}
[scale=1\columnwidth/10cm,font=\normalsize]
\begin{axis}[%
width=8cm,
height=4cm,
at={(0, 0)},
scale only axis,
xmode = log,
xmin=0.25,
xmax=10,
xlabel style={font=\large\color{white!15!black}, yshift=1 ex},
xlabel={Distance},
ymin=-60,
ymax=20,
ylabel style={font=\large\color{white!15!black}, yshift= -1.5 ex},
ylabel={Mismatch Error [dB]},
axis background/.style={fill=white},
xmajorgrids,
ymajorgrids,
]
\addplot [color=black, line width=1.5pt, forget plot]
  table[row sep=crcr]{%
0.1	26.8782363634963\\
0.104761575278966	35.4961384581895\\
0.109749876549306	32.7188402924044\\
0.114975699539774	32.3750220945379\\
0.120450354025878	31.9377162839991\\
0.126185688306602	31.205190982308\\
0.132194114846603	30.8735703603103\\
0.138488637139387	26.9313567730822\\
0.145082877849594	26.4358857463208\\
0.151991108295293	26.037624101313\\
0.159228279334109	25.6750393053617\\
0.166810053720006	25.3478932183172\\
0.174752840000768	25.0121321404959\\
0.183073828029537	24.6722967159969\\
0.191791026167249	24.5439600249697\\
0.200923300256505	24.1793952412416\\
0.210490414451202	23.8012027282211\\
0.220513073990305	23.6475209648278\\
0.231012970008316	23.1919041810011\\
0.242012826479438	11.2872219573995\\
0.253536449397011	10.3715892174571\\
0.265608778294669	9.69659572819546\\
0.278255940220712	9.07485348793415\\
0.291505306282518	8.46061148067821\\
0.305385550883342	7.84395249691067\\
0.319926713779738	7.22355477433814\\
0.335160265093884	6.60072839010439\\
0.351119173421513	5.97625933603496\\
0.367837977182863	5.34909736389466\\
0.385352859371053	4.72249828182822\\
0.403701725859655	4.09284740446134\\
0.42292428743895	3.46096837452001\\
0.443062145758388	2.82585357515342\\
0.464158883361278	2.18619515539888\\
0.486260158006535	1.54303284651115\\
0.509413801481638	0.894839153240298\\
0.533669923120631	0.241057849939182\\
0.559081018251222	-0.418794381050873\\
0.585702081805667	-1.08532175935466\\
0.613590727341317	-1.75676001491064\\
0.642807311728432	-2.43377062354341\\
0.673415065775082	-3.11697443092561\\
0.705480231071864	-3.80377535627764\\
0.739072203352578	-4.49553797149241\\
0.774263682681127	-5.18969968069644\\
0.811130830789687	-5.88562990429097\\
0.849753435908644	-6.5825690427215\\
0.890215085445039	-7.2788854698187\\
0.93260334688322	-7.97387021783834\\
0.977009957299225	-8.66645012992603\\
1.02353102189903	-9.35552073042896\\
1.07226722201032	-10.0404578435371\\
1.12332403297803	-10.7199261228319\\
1.176811952435	-11.3933985868388\\
1.23284673944207	-12.0603444053962\\
1.29154966501488	-12.720043343918\\
1.35304777457981	-13.3715475005762\\
1.41747416292681	-14.0148550531379\\
1.48496826225447	-14.6497029643616\\
1.55567614393047	-15.2754330197361\\
1.62975083462064	-15.8920247058935\\
1.70735264747069	-16.4990934257601\\
1.78864952905743	-17.0966965015454\\
1.87381742286038	-17.6849200763147\\
1.96304065004027	-18.2633148013065\\
2.05651230834865	-18.8327066461921\\
2.15443469003188	-19.3920141067019\\
2.25701971963392	-19.9423372004536\\
2.36448941264541	-20.4835505867277\\
2.47707635599171	-21.0159798445738\\
2.59502421139974	-21.5396811124261\\
2.71858824273294	-22.0549618540418\\
2.8480358684358	-22.5617493077027\\
2.98364724028334	-23.0608872570671\\
3.12571584968824	-23.5523794161874\\
3.27454916287773	-24.0365964087262\\
3.43046928631492	-24.5136911387082\\
3.59381366380463	-24.9840249530159\\
3.76493580679247	-25.4477152685863\\
3.94420605943766	-25.9054164186697\\
4.13201240011534	-26.3569559901368\\
4.32876128108306	-26.8027581391964\\
4.53487850812858	-27.2428752321466\\
4.7508101621028	-27.6777152054774\\
4.97702356433211	-28.1071976689507\\
5.21400828799968	-28.5318586752114\\
5.46227721768434	-28.9511980244415\\
5.72236765935022	-29.3657948102039\\
5.99484250318941	-29.775614357318\\
6.28029144183425	-30.1804727536238\\
6.57933224657568	-30.5804542310051\\
6.8926121043497	-30.9759119640117\\
7.22080901838546	-31.3666627445045\\
7.56463327554629	-31.7523297837408\\
7.92482898353917	-32.1330510312347\\
8.30217568131975	-32.508642894377\\
8.69749002617784	-32.878775685375\\
9.11162756115489	-33.2436332380149\\
9.54548456661834	-33.6026971209282\\
10	-33.9557955846514\\
};
\addplot [color=blue, dashed, line width=1.5pt, forget plot]
  table[row sep=crcr]{%
0.1	-20.2511530892783\\
0.104761575278966	-20.6441136385035\\
0.109749876549306	-21.0374750607481\\
0.114975699539774	-21.431310392327\\
0.120450354025878	-21.8256483584262\\
0.126185688306602	-22.2205429946731\\
0.132194114846603	-22.6159816324064\\
0.138488637139387	-23.0119841690405\\
0.145082877849594	-23.4085235376106\\
0.151991108295293	-23.8055840479224\\
0.159228279334109	-24.2031644532906\\
0.166810053720006	-24.6012004418616\\
0.174752840000768	-24.9997188771616\\
0.183073828029537	-25.3986593996464\\
0.191791026167249	-25.7979992660083\\
0.200923300256505	-26.1976893932842\\
0.210490414451202	-26.5977828873055\\
0.220513073990305	-26.9981591496929\\
0.231012970008316	-27.3988532731649\\
0.242012826479438	-27.7998070613958\\
0.253536449397011	-28.2010380371622\\
0.265608778294669	-28.6025318212835\\
0.278255940220712	-29.0041882294198\\
0.291505306282518	-29.4060873275728\\
0.305385550883342	-29.808156189855\\
0.319926713779738	-30.2103447671267\\
0.335160265093884	-30.612826531136\\
0.351119173421513	-31.0152749029001\\
0.367837977182863	-31.4179382658929\\
0.385352859371053	-31.8206946656038\\
0.403701725859655	-32.2235608116874\\
0.42292428743895	-32.6265255428079\\
0.443062145758388	-33.0295754249599\\
0.464158883361278	-33.4328218555007\\
0.486260158006535	-33.8359942101977\\
0.509413801481638	-34.2393126461957\\
0.533669923120631	-34.642658698409\\
0.559081018251222	-35.0461236884793\\
0.585702081805667	-35.4494679081417\\
0.613590727341317	-35.8531092881428\\
0.642807311728432	-36.2564584571766\\
0.673415065775082	-36.6603999833274\\
0.705480231071864	-37.0640879926848\\
0.739072203352578	-37.4673670610434\\
0.774263682681127	-37.8712601523576\\
0.811130830789687	-38.2745001481503\\
0.849753435908644	-38.678359268148\\
0.890215085445039	-39.0827181093056\\
0.93260334688322	-39.4864147217431\\
0.977009957299225	-39.8900688923367\\
1.02353102189903	-40.2944004955758\\
1.07226722201032	-40.6974352312684\\
1.12332403297803	-41.1011297496834\\
1.176811952435	-41.5057095447837\\
1.23284673944207	-41.9091330420785\\
1.29154966501488	-42.3117801737529\\
1.35304777457981	-42.7162814134402\\
1.41747416292681	-43.1202277713956\\
1.48496826225447	-43.5258454204439\\
1.55567614393047	-43.9274448345985\\
1.62975083462064	-44.3298612221335\\
1.70735264747069	-44.7343616466588\\
1.78864952905743	-45.1394287161653\\
1.87381742286038	-45.5437853169074\\
1.96304065004027	-45.9484080144435\\
2.05651230834865	-46.3509776229977\\
2.15443469003188	-46.7536912192566\\
2.25701971963392	-47.1589221309196\\
2.36448941264541	-47.5627178668134\\
2.47707635599171	-47.9664709893457\\
2.59502421139974	-48.3728448668501\\
2.71858824273294	-48.7699901367099\\
2.8480358684358	-49.176003088956\\
2.98364724028334	-49.5834553777044\\
3.12571584968824	-49.9829739425933\\
3.27454916287773	-50.3836844630813\\
3.43046928631492	-50.788732133377\\
3.59381366380463	-51.1954151291405\\
3.76493580679247	-51.6048912147709\\
3.94420605943766	-51.9947201829698\\
4.13201240011534	-52.3883528975242\\
4.32876128108306	-52.7958406942024\\
4.53487850812858	-53.211903209771\\
4.7508101621028	-53.6056388852973\\
4.97702356433211	-54.0145377164707\\
5.21400828799968	-54.4044891938179\\
5.46227721768434	-54.8078670844915\\
5.72236765935022	-55.2301608285958\\
5.99484250318941	-55.6382823968298\\
6.28029144183425	-56.0093117587255\\
6.57933224657568	-56.417790167434\\
6.8926121043497	-56.8343730397549\\
7.22080901838546	-57.2408002663087\\
7.56463327554629	-57.6340443022837\\
7.92482898353917	-58.0330815220155\\
8.30217568131975	-58.3967107524481\\
8.69749002617784	-58.8443701313786\\
9.11162756115489	-59.2499254777411\\
9.54548456661834	-59.6563305710977\\
10	-60.0061457651333\\
};
\addplot [color=red, dashed, line width=1.5pt, forget plot]
  table[row sep=crcr]{%
0.1	21.20739508664\\
0.104761575278966	20.6077128812684\\
0.109749876549306	20.1158769256972\\
0.114975699539774	22.8400654328215\\
0.120450354025878	21.9928141948309\\
0.126185688306602	21.9249337812814\\
0.132194114846603	21.1487632316003\\
0.138488637139387	20.315872019319\\
0.145082877849594	19.5327187463457\\
0.151991108295293	18.7498604712795\\
0.159228279334109	17.9493569889602\\
0.166810053720006	17.1939987031992\\
0.174752840000768	16.4631100850979\\
0.183073828029537	15.7664703343857\\
0.191791026167249	15.111017113664\\
0.200923300256505	14.4432552108492\\
0.210490414451202	12.4238976758127\\
0.220513073990305	12.0027241172881\\
0.231012970008316	11.5912654101962\\
0.242012826479438	11.0209724009763\\
0.253536449397011	10.3257769187098\\
0.265608778294669	9.64391361260789\\
0.278255940220712	8.97570111423664\\
0.291505306282518	8.31991691488685\\
0.305385550883342	7.67480591076071\\
0.319926713779738	7.03844608301955\\
0.335160265093884	6.40889970273645\\
0.351119173421513	5.78428808044828\\
0.367837977182863	5.16282140655017\\
0.385352859371053	4.54281797171433\\
0.403701725859655	3.92271632881716\\
0.42292428743895	3.3010833469412\\
0.443062145758388	2.67662826596467\\
0.464158883361278	2.04821070872964\\
0.486260158006535	1.41486267817089\\
0.509413801481638	0.775798033955141\\
0.533669923120631	0.130434602332741\\
0.559081018251222	-0.521594390780592\\
0.585702081805667	-1.18043046040879\\
0.613590727341317	-1.84598541051997\\
0.642807311728432	-2.51794540117857\\
0.673415065775082	-3.19579216540512\\
0.705480231071864	-3.87880853718309\\
0.739072203352578	-4.5661249732903\\
0.774263682681127	-5.25674781843827\\
0.811130830789687	-5.94959828181743\\
0.849753435908644	-6.64355068201053\\
0.890215085445039	-7.33746408075859\\
0.93260334688322	-8.03021380964003\\
0.977009957299225	-8.72071911633697\\
1.02353102189903	-9.40795593811197\\
1.07226722201032	-10.0909772607943\\
1.12332403297803	-10.7689103001053\\
1.176811952435	-11.4409757101449\\
1.23284673944207	-12.1064735208809\\
1.29154966501488	-12.7647942318754\\
1.35304777457981	-13.4154131716695\\
1.41747416292681	-14.0578882057284\\
1.48496826225447	-14.6918552178469\\
1.55567614393047	-15.3170200768667\\
1.62975083462064	-15.9331769253603\\
1.70735264747069	-16.5401620378391\\
1.78864952905743	-17.1379014326301\\
1.87381742286038	-17.726361189141\\
1.96304065004027	-18.3055769917605\\
2.05651230834865	-18.8756350297254\\
2.15443469003188	-19.436650617072\\
2.25701971963392	-19.9888190374797\\
2.36448941264541	-20.5323155475838\\
2.47707635599171	-21.0674044856314\\
2.59502421139974	-21.5943578457262\\
2.71858824273294	-22.1134483837572\\
2.8480358684358	-22.6249981286495\\
2.98364724028334	-23.1293281016123\\
3.12571584968824	-23.6267767776689\\
3.27454916287773	-24.1176870917716\\
3.43046928631492	-24.6023758529378\\
3.59381366380463	-25.0812079979666\\
3.76493580679247	-25.5545084661122\\
3.94420605943766	-26.0226273474918\\
4.13201240011534	-26.4858503255036\\
4.32876128108306	-26.9445046225457\\
4.53487850812858	-27.3989204654856\\
4.7508101621028	-27.8493561069623\\
4.97702356433211	-28.2960717299088\\
5.21400828799968	-28.7393590033043\\
5.46227721768434	-29.1794948731815\\
5.72236765935022	-29.6166567210337\\
5.99484250318941	-30.0510966168874\\
6.28029144183425	-30.4830684792379\\
6.57933224657568	-30.9125963538244\\
6.8926121043497	-31.3401311737305\\
7.22080901838546	-31.7656122622259\\
7.56463327554629	-32.1892666015904\\
7.92482898353917	-32.6112835063574\\
8.30217568131975	-33.0317426783982\\
8.69749002617784	-33.4509998883466\\
9.11162756115489	-33.8686168163052\\
9.54548456661834	-34.2852360176462\\
10	-34.7006190035079\\
};
\addplot [color=green, dashed, line width=1.5pt, forget plot]
  table[row sep=crcr]{%
0.1	-41.4416736656709\\
0.104761575278966	-41.691907686279\\
0.109749876549306	-41.9159974906401\\
0.114975699539774	-41.9604239154546\\
0.120450354025878	-41.8730092681865\\
0.126185688306602	-41.6654311734193\\
0.132194114846603	-41.8910054642469\\
0.138488637139387	-41.7604496612798\\
0.145082877849594	-41.8241292109474\\
0.151991108295293	-41.9840985019057\\
0.159228279334109	-41.9388635973633\\
0.166810053720006	-41.9416984709145\\
0.174752840000768	-41.8712522108873\\
0.183073828029537	-41.8692281756336\\
0.191791026167249	-41.9389632613211\\
0.200923300256505	-41.8405973348839\\
0.210490414451202	-41.9468708763176\\
0.220513073990305	-41.9562229441888\\
0.231012970008316	-41.9384385946631\\
0.242012826479438	-41.9695752263957\\
0.253536449397011	-41.9117791137439\\
0.265608778294669	-41.9853175973619\\
0.278255940220712	-41.9605009298175\\
0.291505306282518	-41.9601127897708\\
0.305385550883342	-41.9940090288379\\
0.319926713779738	-41.9972321732341\\
0.335160265093884	-41.9890579804958\\
0.351119173421513	-42.0028603968476\\
0.367837977182863	-41.9933090811513\\
0.385352859371053	-41.9900557972421\\
0.403701725859655	-41.9870645653339\\
0.42292428743895	-41.9986682275825\\
0.443062145758388	-41.9876678284485\\
0.464158883361278	-42.0037283200709\\
0.486260158006535	-42.0032647419537\\
0.509413801481638	-41.9867008858743\\
0.533669923120631	-42.0055649562443\\
0.559081018251222	-42.0006070884731\\
0.585702081805667	-42.0068200584106\\
0.613590727341317	-42.0089999120083\\
0.642807311728432	-42.0078308240307\\
0.673415065775082	-42.0100440073171\\
0.705480231071864	-42.0048437548477\\
0.739072203352578	-42.0076010285545\\
0.774263682681127	-42.0082091685836\\
0.811130830789687	-42.0054934618888\\
0.849753435908644	-42.0098665384775\\
0.890215085445039	-42.0090769509197\\
0.93260334688322	-42.0102382067948\\
0.977009957299225	-42.0086102381247\\
1.02353102189903	-42.0081142024417\\
1.07226722201032	-42.0107216978708\\
1.12332403297803	-42.0073139654878\\
1.176811952435	-42.0105715814162\\
1.23284673944207	-42.01048935151\\
1.29154966501488	-42.0114090515852\\
1.35304777457981	-42.010715567811\\
1.41747416292681	-42.0096775878412\\
1.48496826225447	-42.0113690537748\\
1.55567614393047	-42.0109331803815\\
1.62975083462064	-42.0096895994311\\
1.70735264747069	-42.0103579967314\\
1.78864952905743	-42.010167038589\\
1.87381742286038	-42.0095731299486\\
1.96304065004027	-42.0096073259112\\
2.05651230834865	-42.0109266927371\\
2.15443469003188	-42.0102133408313\\
2.25701971963392	-42.0100701960815\\
2.36448941264541	-42.0104980075073\\
2.47707635599171	-42.0108102738242\\
2.59502421139974	-42.0120183045444\\
2.71858824273294	-42.0103852792063\\
2.8480358684358	-42.0118709199959\\
2.98364724028334	-42.0112614066139\\
3.12571584968824	-42.011585432349\\
3.27454916287773	-42.0104337606242\\
3.43046928631492	-42.0115429425894\\
3.59381366380463	-42.0103300230768\\
3.76493580679247	-42.0108536815621\\
3.94420605943766	-42.0106940107743\\
4.13201240011534	-42.0108267773765\\
4.32876128108306	-42.0103424116257\\
4.53487850812858	-42.0101773087533\\
4.7508101621028	-42.0114760195999\\
4.97702356433211	-42.01094240235\\
5.21400828799968	-42.0112025330303\\
5.46227721768434	-42.0107656013233\\
5.72236765935022	-42.0105618028889\\
5.99484250318941	-42.0110853426157\\
6.28029144183425	-42.0107914174792\\
6.57933224657568	-42.0104587358996\\
6.8926121043497	-42.0105801639784\\
7.22080901838546	-42.0102721028417\\
7.56463327554629	-42.0110049105358\\
7.92482898353917	-42.0108617132779\\
8.30217568131975	-42.0110187749644\\
8.69749002617784	-42.0098971277891\\
9.11162756115489	-42.0103941760927\\
9.54548456661834	-42.0106771278684\\
10	-42.0112198089785\\
};
\addplot [color=blue, only marks, line width=1.5pt, mark=o, mark options={solid, blue}, forget plot]
  table[row sep=crcr]{%
0.1	-20.2511530892783\\
0.120450354025878	-21.8256483584262\\
0.145082877849594	-23.4085235376106\\
0.174752840000768	-24.9997188771616\\
0.210490414451202	-26.5977828873055\\
0.253536449397011	-28.2010380371622\\
0.305385550883342	-29.808156189855\\
0.367837977182863	-31.4179382658929\\
0.443062145758388	-33.0295754249599\\
0.533669923120631	-34.642658698409\\
0.642807311728432	-36.2564584571766\\
0.774263682681127	-37.8712601523576\\
0.93260334688322	-39.4864147217431\\
1.12332403297803	-41.1011297496834\\
1.35304777457981	-42.7162814134402\\
1.62975083462064	-44.3298612221335\\
1.96304065004027	-45.9484080144435\\
2.36448941264541	-47.5627178668134\\
2.8480358684358	-49.176003088956\\
3.43046928631492	-50.788732133377\\
4.13201240011534	-52.3883528975242\\
4.97702356433211	-54.0145377164707\\
5.99484250318941	-55.6382823968298\\
7.22080901838546	-57.2408002663087\\
8.69749002617784	-58.8443701313786\\
};
\addplot [color=red, only marks, line width=1.5pt, mark=square, mark options={solid, red}, forget plot]
  table[row sep=crcr]{%
0.1	21.20739508664\\
0.120450354025878	21.9928141948309\\
0.145082877849594	19.5327187463457\\
0.174752840000768	16.4631100850979\\
0.210490414451202	12.4238976758127\\
0.253536449397011	10.3257769187098\\
0.305385550883342	7.67480591076071\\
0.367837977182863	5.16282140655017\\
0.443062145758388	2.67662826596467\\
0.533669923120631	0.130434602332741\\
0.642807311728432	-2.51794540117857\\
0.774263682681127	-5.25674781843827\\
0.93260334688322	-8.03021380964003\\
1.12332403297803	-10.7689103001053\\
1.35304777457981	-13.4154131716695\\
1.62975083462064	-15.9331769253603\\
1.96304065004027	-18.3055769917605\\
2.36448941264541	-20.5323155475838\\
2.8480358684358	-22.6249981286495\\
3.43046928631492	-24.6023758529378\\
4.13201240011534	-26.4858503255036\\
4.97702356433211	-28.2960717299088\\
5.99484250318941	-30.0510966168874\\
7.22080901838546	-31.7656122622259\\
8.69749002617784	-33.4509998883466\\
};
\addplot [color=green, only marks, line width=1.5pt, mark=diamond, mark options={solid, green}, forget plot]
  table[row sep=crcr]{%
0.1	-41.4416736656709\\
0.120450354025878	-41.8730092681865\\
0.145082877849594	-41.8241292109474\\
0.174752840000768	-41.8712522108873\\
0.210490414451202	-41.9468708763176\\
0.253536449397011	-41.9117791137439\\
0.305385550883342	-41.9940090288379\\
0.367837977182863	-41.9933090811513\\
0.443062145758388	-41.9876678284485\\
0.533669923120631	-42.0055649562443\\
0.642807311728432	-42.0078308240307\\
0.774263682681127	-42.0082091685836\\
0.93260334688322	-42.0102382067948\\
1.12332403297803	-42.0073139654878\\
1.35304777457981	-42.010715567811\\
1.62975083462064	-42.0096895994311\\
1.96304065004027	-42.0096073259112\\
2.36448941264541	-42.0104980075073\\
2.8480358684358	-42.0118709199959\\
3.43046928631492	-42.0115429425894\\
4.13201240011534	-42.0108267773765\\
4.97702356433211	-42.01094240235\\
5.99484250318941	-42.0110853426157\\
7.22080901838546	-42.0102721028417\\
8.69749002617784	-42.0098971277891\\
};
\end{axis}

\end{tikzpicture}%